\documentclass[11pt,a4paper]{article}
\pdfoutput=1
\usepackage{jheppub}
\usepackage[T1]{fontenc}
\usepackage{hyperref}
\usepackage{comment}
\usepackage{outlines}
\usepackage{setspace}
\usepackage{cancel}
\usepackage{stmaryrd}
\usepackage{amsthm}
\usepackage[normalem]{ulem}
\title{On the Charge Algebra of Causal Diamonds in Three Dimensional Gravity}
\author{Pranav Pulakkat}
\affiliation{Maryland Center for Fundamental Physics,\\ University of Maryland, College Park, MD 20740, USA}
\emailAdd{pranavp@umd.edu}
\date{March 2023}
\newcommand{\dS}{\partial\Sigma}
\newcommand{\Np}{$\mathcal{N}^+$}
\newcommand{\Nm}{$\mathcal{N}^-$}
\newcommand{\qw}{\mathcal{Q_{\bar{\omega}}}}
\newcommand{\qa}{\mathcal{Q}_a}

\newcommand{\Ha}{H\'aj\'i\v cek}    

\abstract{Covariant phase space methods are applied to the analysis of a causal diamond in 2+1-dimensional pure Einstein gravity. It is found that the reduced phase space is parametrized by a family of charges with a dual geometrical interpretation: they are geometric observables on the corner of the diamond, and they generate diffeomorphisms. The Poisson brackets among them close into an algebra. Knowledge of the corner charges therefore permits reconstruction of the diamond geometry, which realizes a form of local holography. The results are contrasted with the literature, and the path to a quantum description of spacetime geometry is discussed.}
\begin{document}
\flushbottom
\maketitle
\section{Introduction}
\par
One of the deepest properties of general relativity is general covariance, or the absence of nondynamical background structures beyond the oriented spacetime manifold. This is manifest for the pure gravity Einstein-Hilbert action
\begin{equation}\label{action}
    S=\int (R-2\Lambda)\epsilon,
\end{equation}
where $g$ is the dynamical metric and $\epsilon$ is the associated volume form compatible with the spacetime orientation. Additional matter terms will also be generally covariant as long as they are constructed purely out of the metric and other dynamical fields. It seems reasonable to expect that this property will continue to hold for the true theory of quantum gravity, whatever that may be. In the formalism of field theory on a manifold, this implies that diffeomorphisms supported in the interior are gauge symmetries, in that two configurations related by such transformations cannot be physically distinguished. Observables must be invariant under gauge transformations, so they cannot be localized in finite regions of the bulk spacetime --- if they were, diffeomorphisms would certainly change their values. This complicates the description of local geometry and subsystems in a manner which is amenable to quantum theory. Much work has been devoted to defining relational observables that are anchored to dynamical features of particular states (see for example \cite{dress} and its references).
\par
However, there are some observables that are comparatively straightforward to construct, and have a simple geometrical structure. These are associated with diffeomorphisms that extend to the spacetime boundary and are compatible with boundary conditions, which we will call \textit{surface symmetries}. Such symmetries have been known for a long time \cite{surface_integrals} to \textit{not} necessarily be gauge. Instead they correspond to flows on phase space that have nontrivial generators, such as the ADM Hamiltonian. The covariant phase space formalism makes their analysis especially 
transparent \cite{iyer-wald,lee+wald,subsystem,local,harlow}, and reveals that the generators usually have a simple description as integrals of local densities over a codimension-2 cross-section of the boundary. These observables seem to give a handle on the geometry, at least near the cross-section, and have a straightforward interpretation in terms of the flows that they generate.
\par
In the three dimensional pure gravity case one can hope for more, because solutions have especially simple properties \cite{carlip}. The Einstein equation is
\begin{equation}\label{einstein}
    R_{ab}=2\Lambda g_{ab},
\end{equation}
and the Weyl curvature tensor vanishes, so the Riemann tensor is fully determined by the Ricci part:
\begin{equation}\label{Riemann}
    R_{abcd}=\Lambda [g_{ac}g_{bd}-g_{ad}g_{bc}].
\end{equation}
As a consequence of this, any solution has constant sectional curvature $\Lambda$ and is locally isometric to a maximally symmetric spacetime (MSS).  Thus any linearized perturbation of a solution is generated by some diffeomorphism. Physically, this corresponds to the fact that there are no gravitational waves in three dimensions. All degrees of freedom should be associated with diffeomorphism generators at a codimension-2 surface.
\par
This principle is realized at asymptotically AdS boundaries by the symmetry algebra discovered by Brown and Henneaux \cite{Brown_Henneaux}, the generators of which completely parametrize the pure gravity phase space \cite{Maloney_Witten}. Recent work on 2+1 dimensional causal diamonds \cite{2+1Diamond,rodrigo_classical,rodrigo_quantum} also seems to support this picture by describing and quantizing a reduced phase space for such systems in terms of observables localized to the diamond corner. However, in that work those observables are not clearly interpreted as 
spacetime diffeomorphism charges. Furthermore, the phase space studied
there was that of diamonds with a fixed corner metric, via methods that require $\Lambda\leq0$. To view this 
as a subspace of the larger phase space of all diamonds one must fix
not only the corner metric but also the corner boost freedom (e.g. by 
fixing the tangent plane to the maximal slice at the corner), which 
is a gauge freedom within the fixed corner metric phase space 
but not in the full phase space (see section \ref{subsec:relation} for more details). In the present study we will 
analyze this full phase space, employing covariant methods that work for arbitrary $\Lambda$. Section 2 will define the system and set up the basic technology. In sections \ref{sec:sym} and \ref{sec:frames} we will identify a complete set of surface symmetries, and in section \ref{sec:charge} we will compute the associated charges, which are given by geometric expressions on the corner. Importantly, \ref{subsec:norm} will find charges generating diffeomorphisms that move the diamond corner normal to itself. This is a fundamentally new result, and is the only aspect of the analysis up to this point that requires the spacetime dimension to be three.
\par
 Together, these charges parametrize the reduced phase space. They will be found to form a closed Poisson algebra in section \ref{sec:algebra}; the results are arranged in tables \ref{tab:charges} and \ref{tab:algebra}. By exponentiating the action of a charge on the others, one obtains the transformation of these geometric quantities on the corner under the action of the finite diffeomorphism given by exponentiating the corresponding infinitesimal surface symmetry. Since the algebra includes the generators of transformations that move the corner, this diffeomorphism can be constructed to map an arbitrary loop embedded inside the diamond to the corner. The geometry of the loop can be read off from the values of the corner charges after applying the transformation. Therefore, the algebra allows reconstruction of the complete geometry from the corner charges, which implements a form of the holographic principle. The overcompleteness of the algebra is addressed in section \ref{sec:coadjoint}, where the coadjoint orbit method is applied to find all of its Casimir invariants. The discussion will place these results in the context of prior work and remark on future directions regarding the quantization of the system.

\subsection{Conventions}\label{subsec:conven}
 We will mostly follow the conventions of \cite{harlow}. The metric signature is mostly plus, and the total spacetime dimension is $D$ in general; it will be restricted to three when the particularities of that dimension become relevant, just prior to equation \eqref{3D}. As assumed already in equation \eqref{action}, the units are such that $16\pi G_N =1$. All indices are abstract, and are usually suppressed unless needed for an explicit contraction. Since we are using the covariant phase space formalism, we will be dealing with differential forms on both spacetime and a space of field histories. Both will follow the same conventions for the wedge product and exterior derivative:
\begin{equation}\label{forms}
    \begin{aligned}
        (\alpha\wedge\beta)_{a_1...a_pb_1...b_q} &= \frac{(p+q)!}{p!q!}\alpha_{[a_1...a_p}\beta_{b_1...b_q]}\\
        d\alpha_{ca_1...a_p} &=(p+1)\partial_{[c}a_{a_1...a_p]};
    \end{aligned}
\end{equation}
however the field space wedge product does not appear explicitly in any equations. Exterior differentiation appears on both spacetime and field space. To distinguish between the two, the spacetime exterior derivative is denoted $d$ and the prephase space exterior derivative (also referred to as the \textit{variation}) is denoted $\delta$. Many forms have both field space and spacetime indices which are suppressed; for clarity, insertion of a vector $X$ into the first spacetime index is denoted with a lowercase $i_X$, and insertion of a field space vector (which will also be called a \textit{flow)} $\hat{\chi}$ into the first field space index is denoted with a capital $I_{\hat{\chi}}$.

\section{Covariant Phase Space Preliminaries}\label{sec:cov}
\par
Our object of study is a causal diamond regarded as a system in its own right, without any consideration of a universe outside of it. The first step in the analysis is to identify the \textit{prephase space} of all possible spacetimes consistent with the structure of a diamond, which correspond to states of the system. In principle, such a spacetime is a solution to Einstein's equations with null boundary containing a fixed, spacelike codimension-2 sphere $\mathcal{S}$ called the \textit{corner}. $\mathcal{S}$ is required to bound a closed codimension-1 disk $\Sigma$, such that the entire diamond is the domain of dependence of $\Sigma$. This description presents some difficulties from a practical view, however; not all such diamonds have the same global manifold structure. For one thing, there will be dynamically formed caustics and crossings of null generators on the boundary, which make the differentiable structure of the spacetime at the boundary dependent on the particular solution. In fact it is not even possible to fix the topology. In the case where the cosmological constant is positive, for example, one can consider diamonds that can be embedded in de Sitter. Sufficiently small diamonds will be compact, as the boundary will form a closed surface. However large diamonds, such as a static patch, may not be. These are still admissible in the prephase space, since they are domains of dependence of a finite closed disk.
\par
\begin{figure}
     \centering
     \includegraphics[scale=.7]{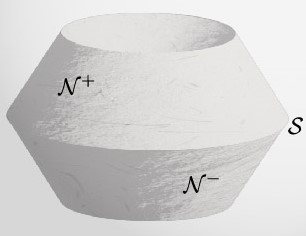}
     \caption{The region $\mathcal{R}$, containing the corner $\mathcal{S}$ and horizons \Np,\;\Nm. Although not shown, $\partial \mathcal{R}$ also contains portions that smoothly cap off the top and bottom of the illustrated boundary.}
     \label{fig:diamond}
 \end{figure}

This poses a problem for the covariant phase space formalism, which works with a prephase space of solutions on a fixed manifold. To get around this, we will describe the solutions only in a compact, ball-shaped region $\mathcal{R}$ (see figure \ref{fig:diamond}) with the following properties:
\begin{itemize}
 \item The intersection of the diamond boundary with $\mathcal{R}$ consists of partial past and future Cauchy horizons \Np\; and \Nm, intersecting transversely at $\mathcal{S}$. Each of these are diffeomorphic to $S^{D-2}\times[0,1]$, with $\mathcal{S}=S^{D-2}\times \{0\}$ for both. In other words they form collars of $\mathcal{S}$ in a compact neighbourhood.
  \item The spacetime is smooth in $\mathcal{R}$, except at the corner $\mathcal{S}$ itself. Derivatives can still be taken at $\mathcal{S}$ by appropriate limits. 
\item For convenient application of Stokes' theorem, we will follow orientation conventions corresponding to those of \cite{harlow}. $\mathcal{S}$ is oriented as the boundary of $\Sigma$, which is itself oriented as the boundary of its past in $\mathcal{R}$. \Np\;and \Nm\; are oriented as pieces of the boundary of $\mathcal{R}$.
\end{itemize}

\par
The covariant phase space treatment will be modeled on $\mathcal{R}$, rather than the full, solution dependent diamond manifold. This has the consequence that although the equations of motion are derived by setting the functional derivative of the action \eqref{action} to 0, we are not imposing a true variational principle that requires the solution to be a stationary point of the action functional in some configuration space. There are two reasons for this: the action is an integral over the \textit{entire diamond manifold}, not just $\mathcal{R}$, which is not a fixed object in this theory; and the stationarity of the action generally requires the imposition of boundary conditions as well as the addition of a boundary term.  In \cite{anomaly,framework,padmanabhan} such combinations of boundary conditions and terms are derived for null boundaries in GR. However, these boundary conditions are unnatural for the system under consideration, which needs no further conditions to generate the spacetime geometry from initial data on the slice $\Sigma$. The only natural restriction on the horizons \Np\;and \Nm\;is that they are null --- anything beyond this would eliminate some freedom in choosing the state of the system.
\par
Despite the absence of a true variational principle, it is still possible to proceed with the covariant phase space formalism. We will take the prephase space $\mathcal{P}$ to consist of all possible smooth metrics on $\mathcal{R}$ satisfying the above properties and vacuum Einstein equation with cosmological constant $\Lambda$. As is standard, we will be cavalier about the topology on this space, only assuming there is a suitable choice which turns it into a Banach \cite{lee+wald}, or more generally Frechet \cite{spaces}, manifold on which the tools of exterior calculus apply. Consult section \ref{subsec:conven} for notation.

The \textit{symplectic potential form} $\theta$ is a spacetime codimension 1 form, prephase space 1-form constructed locally out of fields, variations of fields, and derivatives thereof (as well as background structures), that satisfies the equation $\delta L =d\theta$ on shell. Note that there is an ambiguity in the choice of $\theta$ corresponding to the addition of an arbitrary locally constructed closed (in fact, by the arguments of \cite{Wald_locally_constructed}, locally constructed exact) form to $\theta$, which will integrate out to a corner term. For any given choice the variation $\omega=\delta\theta$ is the \textit{symplectic current}, which upon being integrated over $\Sigma$ yields a prephase space two-form $\Omega$ which is manifestly closed, and thus is a \textit{presymplectic form} on $\mathcal{P}$. It can also be shown that $\omega$ is closed under $d$ on shell, so $\Omega$ is the same for any slice that has the same boundary. Null directions of $\Omega$ are identified with gauge transformations, so the reduced phase space $\mathcal{\tilde{P}}$ is the quotient of $\mathcal{P}$ by the null orbits. Observables are functions on the prephase space that are invariant on these orbits.
\par
The aforementioned ambiguity in the symplectic potential descends to the symplectic form. Different choices can potentially differ in both the class of gauge invariant observables and the flows that these generate. A well known principle for how to select a particular symplectic potential form (or at least restrict the possible choices) originates in \cite{Compere_Marolf_corner} by Compere and Marolf. This was further developed by Harlow and Wu in \cite{harlow} for the case where the spacetime has a timelike boundary, and later extended to null boundaries in \cite{anomaly, framework}. 
These authors argue that the symplectic potential form at the timelike (respectively null) boundary, for given boundary conditions, should be chosen to cancel the variation of a boundary term in the action. Alternatively (without reference to boundary terms), the prescription requires that the flux of \textit{symplectic current} through that boundary vanishes for the chosen conditions, so that different cuts of the boundary yield the same presymplectic form. The perspective is that the system is \textit{closed}, so different choices of slice do not affect the construction of the presymplectic structure. This has no bearing on our problem, since the system is open; information can flow through the horizon and escape the system, so different cuts of the horizons \Np\; or \Nm\; should not be expected to yield equivalent presymplectic forms.  Since the corner $\mathcal{S}$ bounds a Cauchy slice for the diamond, the presymplectic form associated with $\mathcal{S}$ is naturally taken to be the total presymplectic form for the system.
\par
In the absence of such a recipe, the simplest criterion is that the symplectic potential form is constructed covariantly, i.e. involving no background structures, and with the fewest possible number of derivatives, which leads to the following form for the Einstein theory:
\begin{equation}\label{symp}
    \theta=[\nabla^c(\delta g^{ab}g_{ab})-\nabla_b \delta g^{bc}]\epsilon_c.
\end{equation}
Although it is possible to derive an explicit expression for $\Omega$ from this, we will not need this for any subsequent calculations due to an identity discussed in the next section.
\par
In 2+1 dimensions the prephase space has some special properties which will be important. These arise from the fact that the solutions are locally isometric to the MSS of appropriate cosmological constant. For any particular solution, consider a small patch $U$ inside $\mathcal{R}$ which can be embedded in the MSS. It is possible to diffeomorphically contract $\mathcal{R}$ into $U$. Composing the embedding of the patch with the contraction yields an isometric embedding of $\mathcal{R}$ into the MSS\footnote{While it may also be possible to embed into other spacetimes, such as BTZ black holes, the contractibility of $\mathcal{R}$ means that none of the topological information of these spacetimes is captured by solutions in the prephase space.}. This also implies that the complete diamond can be embedded as the domain of dependence in the MSS of the embedded $\mathcal{R}$. Since it is obviously possible to deform any embedding of $\mathcal{R}$ into any other via a smooth homotopy acting on the MSS, this means that given any solution on $\mathcal{R}$, it is possible to deform it to any other by applying a time dependent diffeomorphism flow. In other words, diffeomorphisms \textit{act transitively} on the prephase space. 

\section{Surface Symmetries of General Relativity}\label{sec:sym}
Our goal is to find a set of observable (gauge invariant) charges that generate a set of solution-dependent diffeomorphisms, working for the moment in general dimension. We will eventually specialize to the case of 
 2+1 dimensional gravity,  where we expect to be able to extend this set to be complete, in that its elements will completely parametrize the reduced phase space $\tilde{\mathcal{P}}$. This is due to the fact that \textit{all} tangency directions to prephase space are diffeomorphisms, so any observable is a diffeomorphism generator. If the variations of these charges span the cotangent space at each point of $\tilde{\mathcal{P}}$, then they form a complete parametrization. This is equivalent to the condition that the flows they generate form a basis for the tangent space of $\tilde{\mathcal{P}}$, which is the vector space of linearized diffeomorphisms modulo gauge transformations.
\par
As discussed in the introduction, general covariance implies that inserting a diffeomorphism flow into the presymplectic form results in a prephase space 1-form localized to the boundary $\mathcal{S}$, often referred to as a generalized "charge variation" despite that a priori it may not be possible to integrate to obtain an actual charge. Explicitly we have the following master equation \cite{framework}, often referred to as the Iyer-Wald identity \cite{iyer-wald} (although the original source did not account for solution dependence), which will form the basis for all subsequent computations:
\begin{equation}\label{charge}
    -I_{\hat{\xi}} \Omega = \int_{\dS} \delta Q_\xi -Q_{\delta\xi}-i_\xi\theta.
\end{equation}
Here $\xi$ is the spacetime diffeomorphism vector field, which is a function on prephase space, and $\hat{\xi}$ is the prephase space flow it generates; $Q_\xi$ is the Noether charge form and $Q_{\delta\xi}$ is the same with $\xi$ replaced with its prephase space exterior derivative $\delta\xi$; and $i_\xi \theta$ is the insertion of $\xi$ into the first spacetime index of $\theta$. The first two terms can be reinterpreted as the variation of $Q$ while holding the vector $\xi$ constant with respect to $\mathcal{M}$. The Noether charge form for general relativity is \cite{harlow}
\begin{equation}\label{Noether}
    Q_\xi=-\star d\xi_\flat=-\epsilon^a_{\;\;b} \nabla_a\xi^b.
\end{equation}
This depends only on the value of $\xi$ at $\dS$ and its first derivatives there, and the same is true of the other terms in \eqref{charge}. This yields \textit{a priori} $3D$ independent parameters per point of $\dS$ that classify all physical symmetries; $D$ from $\xi|_{\dS}$, and $2D$ from the remaining unconstrained derivatives, which can be associated with diffeomorphisms that fix all points of $\dS$. Two of the latter (associated with diffeomorphisms that tilt the horizons) are eliminated by the condition that the  horizons remain null, leaving $2D-2$ symmetry parameters. All of these symmetry generators can be described to the relevant order by specifying $\xi$ at the corner and the flow it induces on the normal plane at each point thereof. This allows the following geometric classification:
\begin{outline}
    \1 A two parameter (per point) subalgebra of \emph{superscaling} diffeomorphisms that vanish on the corner, and preserve the normal plane. These can be further divided into two one parameter algebras:
        \2 \textit{Boosts}, which preserve the metric on the normal plane
        \2 \textit{Scalings}, which uniformly scale all vectors in the normal plane.
    \1 A $2(D-2)$ dimensional algebra of \textit{shearing} transformations that also vanish on the corner and Lie flow the normal plane orthogonal to itself. A natural way to specify a particular shearing transformation up to first order off the corner is to choose a pair of vector fields that form a basis of the normal plane at each point of the corner, and then fix the Lie derivative of each vector field under $\xi$ to be a particular vector tangent to the corner. Note that the extension of these vector fields off of the corner does not play any role in this Lie bracket.
\0 These form the Lie algebra of the little group of $\dS$. The full symmetry algebra also includes generators with $\xi$ nonvanishing on the corner. These can be decomposed into:

    \1 Generators with $\xi$ tangent to the corner, which we call \textit{superrotations}. There is considerable multiplicity for any given restriction of $\xi$ to $\dS$, because one can add any generator from the little group Lie algebra without affecting it. To pick out a particular "pure" diffeomorphism generator, we can again employ a pair of basis vector fields for the normal plane at the corner. Requiring that the Lie flow preserves these vector fields determines a unique extension of $\xi$ off the corner to first order, faithfully yielding $D-2$ generators per point.  Again, the extension of the basis vectors off the corner is irrelevant.
    \1 Generators with $\xi$ normal to the corner, or \textit{normal translations}. These can be specified in the same way; fixing their components in the chosen basis and requiring that they leave this basis invariant under Lie flow. Here, the nature of the extension off of the corner \textit{does} matter, since $\xi$ is no longer tangent to it. For any choice of extension, this prescription specifies $2$ generators per point. 
\end{outline}
In this description we've made reference to two structures on the corner; the aforementioned basis vectors for the normal plane, and an implicit labelling of the points on $\mathcal{S}$. The latter allows us to define symmetry parameters as functions or vector fields on $\mathcal{S}$. A more concrete way of looking at this is that we are using a map $\mathcal{S}\rightarrow S^{D-2}$ to label the points of the corner with a reference sphere, which we call a set of "corner coordinates" for short. This allows one to write down smooth functions taking these points to real numbers, or tangent vector fields. The classification of surface symmetries given above is in terms of some set of corner coordinates. In principle it is possible to generalize this to a set of corner coordinates that is dependent on the solution, just as the normal vector fields must be. However, it is simpler to just set the corner coordinates to be a fixed map with respect to the manifold structure of $\mathcal{R}$. This means that we will not have to refer to them explicitly at all, and can mostly just carry out computations with respect to the background manifold structure.
\par
The next step is to apply equation \eqref{charge} to compute the charge variations for each class of transformation. These will be organized in a natural hierarchy, depending on how much structure is required to specify them. To guarantee that they are integrable, the solution dependence of the structures will have to be fixed in a certain way. This will be nontrivial, since the preferred basis for the normal bundle \textit{must} be solution dependent because the normal plane itself varies over solution space. To prepare for the charge calculations, we will take an interval to consider these in more detail.
\section{Scaling Frames}\label{sec:frames}
First note that we can use the null boundary conditions to fix the normal basis at the corner up to a single parameter, by requiring that the two basis vectors $n$ and $l$ at each point lie along the null horizons \Np\;and \Nm\;respectively (and are therefore null normals to $\dS$), and $n\cdot l=1$. The only remaining freedom to vary the basis consists of a scaling of $n$ accompanied with the corresponding reciprocal scaling of $l$. Following \cite{anomaly} we refer to any particular choice of basis (at every point, and for each solution) as a \textit{scaling frame}, although unlike them we do not yet constrain the possible solution dependence.
\par
\begin{figure}
     \centering
     \includegraphics[scale=.4]{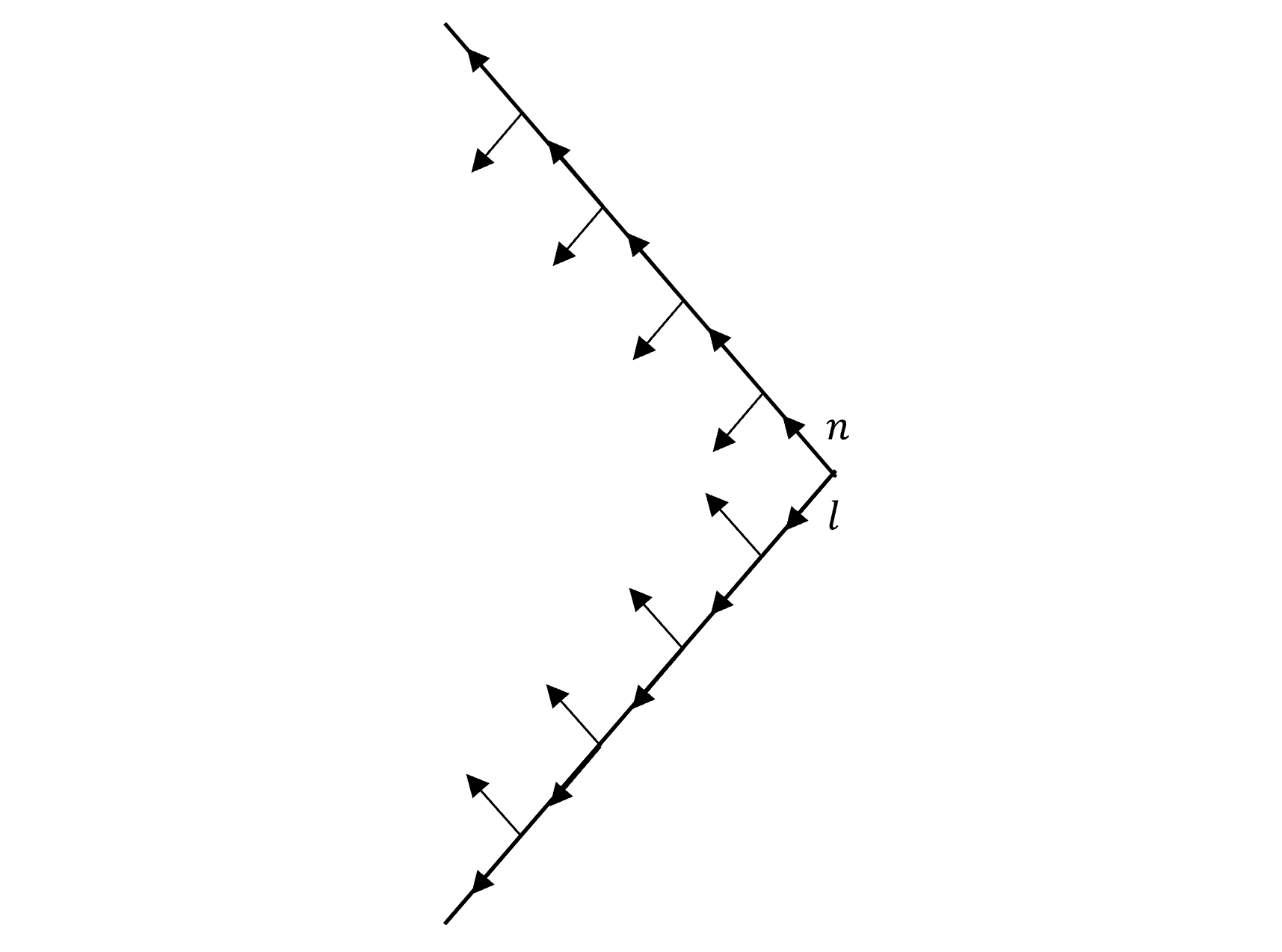}
     \caption{The scaling frame as extended along the two horizons, depicted for a cross-section given by geodesics extending in each direction from a point on the corner.}
     \label{fig:frame}
 \end{figure}

From the discussion in the previous section, we know that the extension of the scaling frame to first order off the corner will be relevant for the normal translation generators. Any scaling frame can be extended along the horizons via parallel transport, setting $\nabla_n n=\nabla_n l=\nabla_l n=\nabla_l l =0$ on \Np\;and \Nm. This also implies that $n$ and $l$ are null and have inner product 1 everywhere on the horizons. For now the extension away from the horizons is left arbitrary. Consult figure \ref{fig:frame}.

Define a projector $q$ onto the tangent subspace orthogonal to both $n$ and $l$ by 
\begin{equation}
q^a_b=\delta^a_b-n^a l_b-l^a n_b
\end{equation}
 The projection of $\nabla_a n^b$ by $q$ can be decomposed  as \begin{equation}
q_a^c\nabla_c n^b  = K_a^b+\bar{\omega}_a n^b; \qquad K_a^b=q_a^c q_d^b\nabla_c n^d \qquad \bar{\omega}_a=q_a^c l_d\nabla_c n^d
\end{equation}
because $n\cdot\nabla n=0$. $K$ is an extrinsic curvature tensor on the horizons associated with the vector $n$, and $\bar{\omega}$ is the \Ha\;1-form. $K$ can be further decomposed into trace and traceless parts:
\begin{equation}
    K^a_b=\frac{1}{D-2}\theta q^a_b+\sigma^a_b
\end{equation}
where $\sigma$ is the traceless shear tensor and $\theta$ can be shown to be the expansion of a codimension-2 surface orthogonal to $n$ and $l$ under the flow of $n$. An identical decomposition can be done for the projected covariant derivative of $l$, so there are actually two shears and expansions (when necessary we will label these with subscripts $+$ and $-$ respectively). However
\begin{equation}\label{Ha_equality}
q_a^c l_d\nabla_c n^d+q_a^c n_d\nabla_c l^d=q_a^c\nabla_c(n\cdot l)=0
\end{equation}
so the \Ha \;1-form for $l$ is the same as that for $n$ with opposite sign, and does not need to be separately defined. When $D=3$, $q$ projects onto a 1-dimensional subspace, so the shear tensor vanishes and the extrinsic curvature tensor is determined purely by the expansion. On the corner, these projections are equal to the full covariant derivatives of $n$ and $l$, since the covariant derivatives of $n$ and $l$ along each other vanish.
\par
Finally it is useful to rewrite the volume form $\epsilon$ in terms of the scaling frame. There is a $(D-2)$-form $\mu=i_li_n\epsilon$ defined on \Np\;and \Nm\;that computes the area of cross-sections of the horizon, which is  positive definite on $\mathcal{S}$ due to the choice of orientation of $\mathcal{S}$. Note that $\mu$ is \textit{independent} of the scaling frame. We have
\begin{equation}\label{volume}
    \epsilon=l\wedge n\wedge \mu
\end{equation}
where $l$ and $n$ are in lowered form. Using this expression, the pullback of the Noether charge form \eqref{Noether} to $\mathcal{S}$ can be rewritten as 
\begin{equation}\label{Noether_Frame}
    Q_\xi=-\epsilon^a_{\;\;b}\nabla_a\xi^b =\nabla_a\xi^b(n^a l_b-l^a n_b)\mu.
\end{equation}
\par 
Scaling frames are extremely useful for computing all diffeomorphism charges, and are crucial for defining most of them, as argued in section \ref{sec:sym}. At first, we will work with a scaling frame that can depend on the solution in arbitrary way. However, from subsection \ref{subsec:diffeo} onwards we will need to fix the solution dependence of the scaling frame in a particular way, which turns out to be that it transforms covariantly under all symmetries \textit{except} the boosts. This condition will arise from requiring that the charge variations associated with superrotations and normal translations are integrable.
\section{Charge Computations}\label{sec:charge}
\subsection{Corner Little Group}\label{subsec:little}
We begin computing charges according to the classification given in section \ref{sec:sym}.  The boost symmetry is the simplest, so it will be used as a model for the techniques used in subsequent calculations.The relevant input to equation \eqref{charge} consists of the generating vector at the corner and its first derivatives. Recall that we have specified the latter in terms of the action of the symmetry on the normal plane to the corner, given by the Lie brackets with $n$ and $l$. The generator is therefore specified to first order by 
\begin{equation}\label{boost_diffeo}
   [\xi,n]=\lambda n\qquad[\xi,l]=-\lambda l \qquad \xi|_{\dS} = 0,
\end{equation}
 where  the boost parameter $\lambda$ is a solution independent scalar function on the corner. As discussed previously, this transformation is independent of the particular choice of scaling frame used to describe it. The Noether charge form on $\dS$ is 
\begin{equation}\label{boost charge}
\begin{aligned}
    Q_\xi&=\nabla_a\xi^b(n^a l_b-l^a n_b)\mu\\
         &=[(\cancel{\nabla_\xi n}-[\xi,n])\cdot l-(\cancel{\nabla_\xi l} - [\xi,l])\cdot n]\mu \\
         &=[-\lambda n\cdot l -\lambda l\cdot n]\mu \\
         &=-2\lambda\mu\\
\end{aligned}
\end{equation}
where the first line uses equation \eqref{Noether_Frame}. The main idea of the calculation is to rewrite the covariant derivatives of $\xi$ on the first line in terms of Lie derivatives and covariant derivatives of $n,l$. The former are given by the definition of the boost symmetry vector, and the latter are geometric quantities studied in section \ref{sec:frames}. All that remains is to simplify the resulting expression. The term $i_\xi\theta$ clearly pulls back to 0 on $\dS$, so the only potential obstruction to the integrability of the charge variation \eqref{charge} is 
\begin{equation}\label{boost integrability}
\begin{aligned}
    Q_{\delta\xi}&=[(\cancel{\nabla_{\delta\xi} n}-[\delta\xi,n])\cdot l-n\leftrightarrow l]\mu\\
                 &=[(-\delta([\xi,n])+[\xi,\delta n])\cdot l- n\leftrightarrow l ]\mu\\
                 &=[(-\lambda\delta n+[\xi,(\delta n\cdot l)n+\cancel{(\delta n\cdot n)} l + q\delta n])\cdot l-n\leftrightarrow l]\mu\\
                 &=[(\cancel{-\lambda\delta n\cdot l}+\cancel{\nabla_\xi}(\delta n\cdot l)(n\cdot l)+\cancel{(\delta n\cdot l)\lambda (n\cdot l)}+\cancel{[\xi,q\delta n]}-n\leftrightarrow l]\mu\\
                 &=0.
\end{aligned}
\end{equation}
The first equality is obtained from the same technique used in \eqref{boost charge} and the observation that $\delta\xi$ vanishes on the corner. The second is obtained by integrating a variation inside a Lie bracket by parts in such a way as to remove all variations from $\xi$, which is convenient because $\xi$ itself has a rather implicit solution dependence. This can be done because the definition of the Lie derivative does not involve any structures that vary. The third is obtained by decomposing $\delta n$ into its projection tangential and normal to the corner. The remaining steps are straightforward cancellations using the properties of the scaling frame and the fact that $\xi$ is 0 on the corner. In the final expression all terms cancel or vanish, showing that the integral of the Noether charge \eqref{boost charge} is an observable which generates the boost symmetry. Note that this argument relied on $\lambda$ being a solution independent constant at each point of the corner. Another way of looking at this is that $\lambda$ is implicitly specified as a function of the fixed set of corner coordinates, which will also be the case for all subsequent diffeomorphism parameters.  There is an exception when $\lambda$ is \textit{uniform}, in which case the transformation and total charge are determined without needing reference to any background structures on $\mathcal{S}$ at all.
\par
These calculations \textit{mutatis mutandis} show that the Noether charge for the scaling symmetry, which has 
\begin{equation}\label{scaling_diffeo}
   [\xi,n]=\lambda n\qquad[\xi,l]=\lambda l \qquad \xi|_{\dS} = 0
\end{equation}
 is zero, and that its integral generates the symmetry. In other words scaling transformations are \textit{gauge}.
\par
The same turns out to be true for shearing transformations. Unlike the scaling and boost transformations, however, these require the scaling frame to be specified; the \textit{class} of shearings is independent of the choice of scaling frame, but picking out any particular shearing transformation requires setting $[\xi,n]$ and $[\xi,l]$ to be fixed vectors tangent to the corner as well as setting $\xi$ at the corner to vanish:
\begin{equation} \label{shearing_diffeo}
[\xi,n]=q[\xi,n]\qquad [\xi,l]=q[\xi,l] \qquad \delta([\xi,n])=\delta([\xi,l])=0 \qquad \xi|_{\dS} = 0.
\end{equation}
A different scaling frame, with the same choice of tangent vectors, defines a different shearing transformation. The Noether charge is
\begin{equation}\label{shearing charge}
    Q_\xi=[(\cancel{\nabla_\xi n}-[\xi,n])\cdot l-n\leftrightarrow l]\mu=0
\end{equation} 
because $[\xi,n]\cdot l=[\xi,l]\cdot n=0$. The obstruction to integrability is again
\begin{equation}\label{shearing integrability}
    \begin{aligned}
    Q_{\delta\xi}&=[(\cancel{\nabla_{\delta\xi} n}-[\delta\xi,n])\cdot l-n\leftrightarrow l]\mu\\
                 &=[(-\cancel{\delta([\xi,n]})+[\xi,\delta n])\cdot l- n\leftrightarrow l ]\mu\\
                 &=[[\xi,(\delta n\cdot l)n+q\delta n])\cdot l-n\leftrightarrow l]\mu\\
                 &=[(\cancel{\nabla_\xi(\delta n\cdot l)}n+(\delta n\cdot l)[\xi,n]+\cancel{[\xi,q\delta n]})\cdot l-n\leftrightarrow l]\mu\\
                 &=[(\delta n\cdot l)\cancel{[\xi,n]\cdot l}-n\leftrightarrow l]\\
                 &=0
    \end{aligned}
\end{equation}
which verifies that the shearing transformations are pure gauge as well.

\subsection{Superrotations}\label{subsec:diffeo}
The superrotations can also be specified using the scaling frame, setting $\xi$ to a fixed tangent vector to $\dS$ and requiring that the Lie flow preserves the frame:
\begin{equation}\label{diffeo_diffeo}
[\xi,n]=[\xi,l]=0 \qquad \delta\xi|_{\dS}=0
\end{equation}
The Noether charge is 
\begin{equation}\label{superrotation charge}
    \begin{aligned}
        Q_\xi&=[(\nabla_\xi n-\cancel{[\xi,n]})\cdot l-n\leftrightarrow l]\mu \\
             &=2\bar{\omega}(\xi)\mu
    \end{aligned}
\end{equation}
using the fact that $\xi$ is tangent to the corner, so $\nabla_\xi n\cdot l=\bar{\omega}(\xi)$, and that the corresponding term swapping $n$ and $l$ involves the \Ha\: 1-form for \Nm, which differs from $\bar{\omega}$ only by a sign. Since $\xi$ is tangent to the corner, the pullback of $i_\xi\theta$ to the corner is still 0. The obstruction to integrability in \eqref{charge} is again just
 \begin{equation}
     \begin{aligned}
         Q_{\delta\xi}&=[(\cancel{\nabla_{\delta\xi} n}-[\delta\xi,n])\cdot l-n\leftrightarrow l]\mu\\
                 &=[(-\cancel{\delta([\xi,n])}+[\xi,\delta n])\cdot l- n\leftrightarrow l ]\mu\\
                 &=[([\xi,(\delta n\cdot l)n+q\delta n])\cdot l-n\leftrightarrow l]\mu\\
                 &=[(\nabla_\xi(\delta n\cdot l)n\cdot l+(\delta n\cdot l)\cancel{[\xi,n]}\cdot l + (\delta n\cdot l)\cancel{[\xi,q\delta n]\cdot l})-n\leftrightarrow l]\mu\\
                 &=[\nabla_\xi (\delta n\cdot l-\delta l\cdot n)]\mu
     \end{aligned}
 \end{equation}
 \par
 For this  to vanish, it is natural to impose that 
 \begin{equation}\label{frame_dependence}
 \delta n\cdot l-\delta l \cdot n =0
 \end{equation}
 .
 This amounts to requiring that the component of $\delta n$ along $n$ is equal to the component of $\delta l$ along $l$. Supposing that the scaling frame is affected only by flows that affect the metric tensor at points along the corner, all such changes can be induced by diffeomorphisms that vanish at the corner. Therefore we only need to guarantee that \eqref{frame_dependence} holds for the little group symmetries. It is manifestly true for scaling and shearing symmetries, so the only possible variations which can violate it are boosts. We can impose that the scaling frames (i.e. the vectors $n$ and $l$) are \textit{not affected} by boost transformations, but transform covariantly under other diffeomorphisms. 
 \par
 While this guarantees the vanishing of $Q_{\delta\xi}$ above, it is not immediately clear that such a scaling frame can be consistently constructed. To see that it can, we leverage the fact that the scaling and shearing transformations are gauge to first \textit{gauge fix} the normal plane and metric. This is equivalent to fixing $g_{ac}-q^b_cg_{ab}$, or the covectors $n_a$ and $l_a$, at every point on the corner. Now requiring that the frame is invariant under boosts amounts to simply setting $n$ and $l$ equal to fixed vectors tangent to the respective horizons. This guarantees that $\delta n$ and $\delta l$ vanish identically at the corner. Such a gauge fixing is consistent with the definition of the superrotation generator, which fixes $[\xi,n]$ and $[\xi, l]$ to be zero (how it acts away from the corner is irrelevant). This is much more convenient to work with than an arbitrary boost noncovariant scaling frame, so we will continue to use it from here on out.

\subsection{Normal Translations}\label{subsec:norm}
\par We can define the normal translations along \Np\; by choosing a scalar function $\gamma_n$ of the corner, extending it by parallel transport along the geodesics generated by $n$ and $l$ on the horizons, and defining 
\begin{equation}\label{norm_diffeo}
\xi=\gamma_n n
\end{equation}
everywhere.  Analogously we have the symmetry given by a flow $\gamma_l l$. Recall that $n$ and $l$ were extended by parallel transport, so the first derivatives are fixed by this equation without requiring any additional Lie bracket conditions. In fact $[n,l]=\nabla_n l=\nabla_l n=0$, so the gauge fixed scaling frame has zero Lie bracket with either of these vector fields. However, they do not actually preserve the gauge fixing condition, since in general they will affect the metric in such a way as to change the normal plane. This can be solved by supplementing with a shearing transformation $\eta$, so the combined diffemorphism vector is $\xi+\eta$ and the flow is $\hat{\xi}+\hat{\eta}$. Upon insertion into the presymplectic form, $-I_{\hat{\xi}+\hat{\eta}}\Omega=-I_{\hat\xi}\Omega$ since shearing flows have already been shown to be gauge. As long as we suppose that any flow inserted into this prephase space 1-form preserves the gauge fixing conditions, we are free to assume the properties of the gauge fixing hold when evaluating the charge variation according to equation \eqref{charge}. We will do this for $\gamma_n n$, and find that in 2+1 dimensions this is integrable, completing the set of charges without introducing any additional reference structures.
\par

\begin{figure}
    \centering
    \includegraphics[scale=.6]{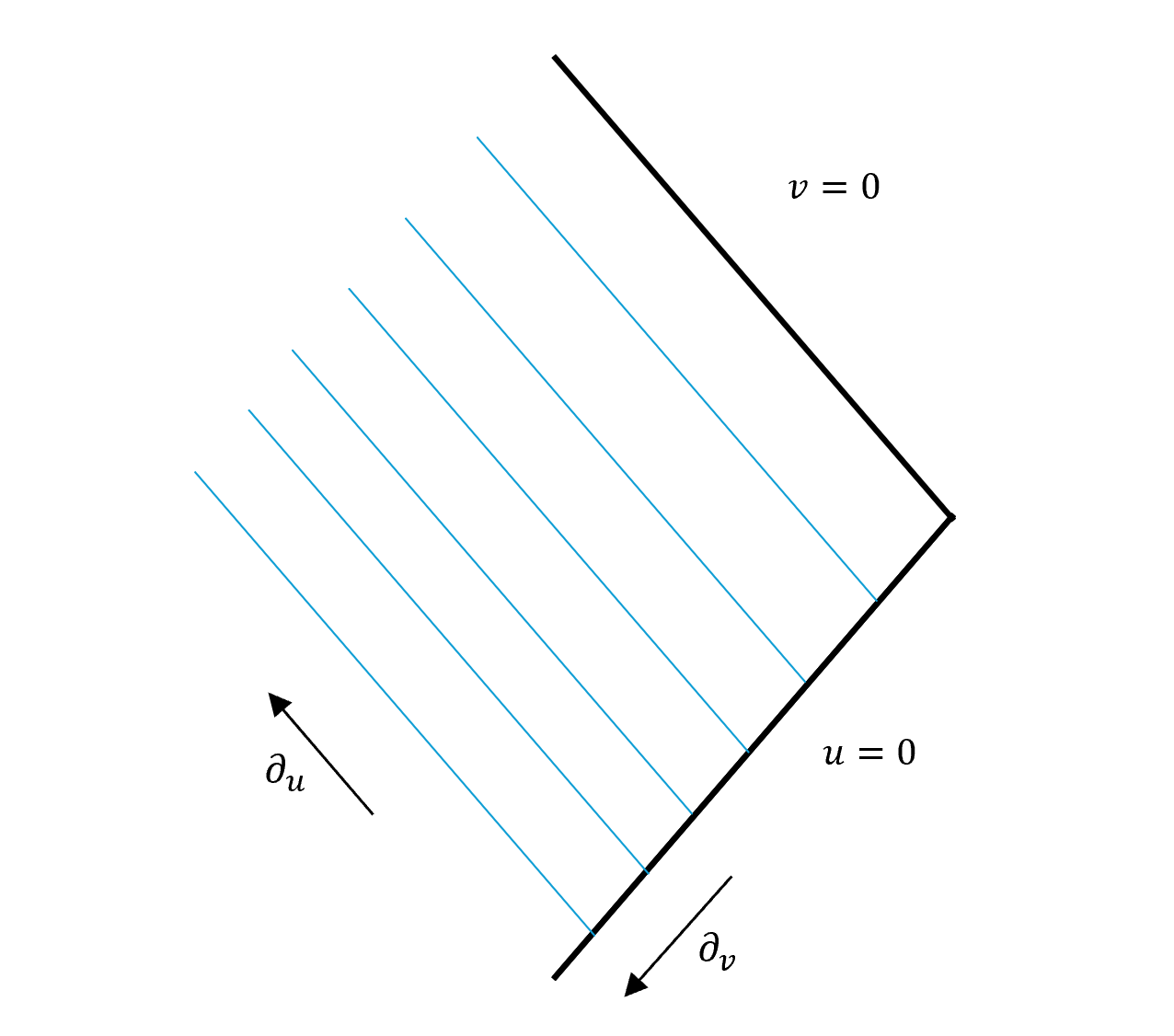}
    \caption{Gaussian null coordinates at a cross-section of fixed angular coordinate $\phi$. The bold black line below is a geodesic on \Nm\; generated by $l$; along this line, $u=0$ and $v$ increases as the affine parameter, such that $\partial_v=l$. The blue lines and upper black line are geodesics generated by $n$, along which $v$ is constant and $u$ increases as the affine parameter going away from \Nm. The upper black line, where $v=0$, lies along \Np.}
    \label{fig:GNC}
\end{figure}

The gauge fixing employed so far does not guarantee that $n$ and $l$ are fixed away from the corner. It will be computationally convenient (though not necessary) to strengthen the gauge fixing by applying a diffeomorphism which is trivial to first order at the corner, to guarantee that $n$ is constant near $\mathcal{S}$. To do this, take the vector field $n$ on \Nm\;and extend into the bulk via the exponential map, then extend $l$ by parallel transport along the geodesics generated by $n$. This is consistent with the previous definition of $n$ and $l$ on \Np. Now construct coordinates $(u,v,\phi)$ in a neighbourhood of the horizons, where $u,v=0$ and $\phi$ fixed on the corner.  Extend these along \Nm\;by holding $u$ and $\phi$ constant along the null generators $l$ and setting $v$ equal to the affine parameter associated with $l$. Then extend along the integral curves of $n$ by taking $u$ to be the affine parameter associated with $n$ and holding $\phi,v$ constant. Clearly $\partial_u = n$ everywhere, and $\partial_v=l$ on \Nm. This forms a \textit{Gaussian null coordinate} system in a neighbourhood of \Nm\;and \Np\;(depicted in figure \ref{fig:GNC}), as originally introduced in \cite{cosmo_cauchy} -- see \cite{padmanabhan,symm_null,symm_causal} for some related applications.

\par
It is always possible to find a diffeomorphism in some neighbourhood of the corner which maps the functions $(u,v,\phi)$ to some fixed coordinates.  This fixes $n$ everywhere in this open set, as well as $l$ on \Nm\;only. This diffeomorphism does not move the points on the corner, and does not move $n$ or $l$ either, so it is trivial up to first order and therefore gauge. We are thereby justified in gauge fixing $(u,v,\phi)$ to be fixed, which also fixes $n$ in a neighbourhood of $\mathcal{S}$ and $l$ on \Nm\;only. Note that the components of $n$ in these coordinates are constant in this neighbourhood, as are the components of $l$ along \Nm.  Since $[n,l]^a=\partial_u l^a-\cancel{\partial_v n^a}=0\implies\partial_u l^a$ at the corner, all derivatives of $n$ and $l$ vanish to first order off the corner. A useful consequence of this is that the corresponding covariant derivatives of their variations vanish for all solutions, i.e. $\nabla_{n,l}\delta(n,l)=0$. This gauge-fixing will be referred to as the $n$-adapted fixing, since it fixes $n$; there is a distinct construction for an $l$-adapted fixing with exactly analogous properties by shooting geodesics from \Np, which is equally useful.

\par
The charge variation for the translations exhibits a different structure from the previous symmetries.  The Noether charge vanishes:
\begin{equation}
    \begin{aligned}
        Q_\xi&=[\nabla_n \xi\cdot l-\nabla_l \xi\cdot n]\mu \\
             &=\gamma_n[ \cancel{\nabla_n n}\cdot l-\cancel{\nabla_l n}\cdot n]\mu\\
             &=0.
    \end{aligned}
\end{equation}
Employing the $n$-adapted system for convenience, $Q_{\delta\xi}$ also vanishes:
\begin{equation}
    \begin{aligned}
         Q_{\delta\xi}&=[\nabla_n \delta\xi\cdot l-\nabla_l \delta\xi\cdot n]\mu\\
         &=\gamma_n[\nabla_n \cancel{\delta n}\cdot l-\nabla_l \cancel{\delta n}\cdot n]\mu\\
         &=0.
    \end{aligned}
\end{equation}
For the first time, however, $i_\xi\theta_R$ \textit{does} contribute at the corner. The pullback to $\mathcal{S}$ can be computed from the symplectic potential decompositions found in \cite{anomaly, brown-york}, but in the interest of being more self-contained we will work it out from scratch in the $n$-adapted system here:
\begin{equation}\label{insert}
    \begin{aligned}
        i_\xi\theta_R ^* = &[\nabla^a(\delta g^{cd}g_{cd})-\nabla_c\delta g^{ac}]\gamma_n n^b[l_a n_b-n_a l_b]\mu\\
                      &=\gamma_n[n_a\nabla_c\delta g^{ac}-\nabla_n (\delta g^{cd}g_{cd})]\mu\\
                      &=\gamma_n[\nabla_c(n_a\delta g^{ac})-\nabla_a n_c \delta g^{ac}-\nabla_n (\delta g^{cd}g_{cd})]\mu.\\
    \end{aligned}
\end{equation}
The first term can be computed by writing $\nabla_c(n_a\delta g^{ac})=-\nabla^c(n^a\delta g_{ac})=-\nabla^c(\delta n_c)$ and expanding the contraction using a frame consisting of $n$, $l$, and a basis of vectors $\chi^{(i)}$ orthogonal to them. The corresponding coframe consists of $n_a$ (dual to $l$), $l$ (dual to $n$), and a basis of 1-forms which annihilate $n$ and $l$, dual to (and also denoted by) $\chi$. Carrying out this computation,
\begin{equation}
    \begin{aligned}
      -\nabla^c(\delta n_c)&=-l^c\nabla_n(\delta n_c)-n^c\nabla_l(\delta n_c)-\sum_i  \chi^c\nabla_\chi (\delta n_c)\\
      &=\cancel{\nabla_n l^c} \delta n_c -\nabla_n(\cancel{\delta(n\cdot l)}-n\cdot\delta l)+\cancel{\nabla_n n^c} \delta n_c-\nabla_l(\cancel{\delta(n\cdot n)})-\sum_i  \chi^c\cancel{\nabla_\chi (\delta n_c)}\\
      &=\cancel{\nabla_n \delta l}\cdot n\\
      &=0
    \end{aligned}
\end{equation}
where we liberally employ the properties of the gauge fixing as well as the fact that $\chi$ is tangent to the corner, so $\nabla_\chi (\delta n_a)$ vanishes since $\delta n_a=0$ on the corner. The last equality follows because of the aforementioned fact that $l$ is fixed up to first order off the corner. \\
Altogether this means that $\nabla_c(n_a\delta g^{ac})=0$. The second term in \eqref{insert} can be decomposed as in section \ref{sec:frames}:
\begin{equation}
\begin{aligned}
-\nabla_a n_c\delta g^{ac}=
&-[K_{ab}+\bar{\omega}_a n_c] \delta g^{ac}\\
&=-[\theta_+ q_a^bg_{bc}+\sigma_{ac}+n_c q^e_a l_d\nabla_e n^d]\delta g^{ac}\\
&=-[\theta_+ q_{ac}+\sigma_{ac}]\delta g^{ac}+\sum_i\cancel{ n_c\chi_a\delta g^{ac}}l_d\nabla_\chi n^d
\end{aligned}
\end{equation}
where $q_{ac}$ is the metric projected by $q$ (whether on one or both indices doesn't matter). 
\par
From here on out we will set $D=3$, upon which the shear vanishes, so on the corner this reduces to
\begin{equation}\label{3D}
-\nabla_a n_c\delta g^{ac}=-\theta_+ q_{ac} \delta q^{ac}=2\theta_+ \frac{\delta\sqrt{|q|}}{\sqrt{|q|}} 
\end{equation}
where $\delta q^{ac}=q^a_b q^c_d \delta g^{bd}=\delta (q^a_b q^c_d g^{bd})$ is the variation of a partial inverse of $q_{ac}$ on the tangent space to the corner. The last equality follows from the formula for the variation of the determinant $|q|$ of $q_{ab}$. Note that the codegree-2 forms at any point which annihilate $n$ and $l$ form a 1-dimensional vector space, so $\delta \mu$ is proportional to $\mu$ by the logarithmic derivative of $\sqrt{|q|}$. We will employ a simple but suggestive abuse of notation and rewrite this term as 
\begin{equation}\label{expansion term}
-\nabla_an_c\delta g^{ac}=2\theta _+ \frac{\delta \mu}{\mu}.
\end{equation}
The last term in \eqref{insert} can be rewritten in a similar way as 
\begin{equation}\label{third term}
    -\nabla_n(\delta g^{cd} g_{cd})=2\nabla_n(\frac{\delta\epsilon}{\epsilon}).
\end{equation}
The variation $\delta\mu$ can be computed in terms of $\delta\epsilon$ anywhere on the horizons from its definition:
\begin{equation}\label{dmu}
\begin{aligned}
    \delta\mu = &[i_{\delta l}i_n+i_l \cancel{i_{\delta _n}}]\epsilon+ i_l i_n\delta\epsilon\\
    &=[(n\cdot l)(\delta l\cdot n)-\cancel{(n\cdot n)}(\delta l\cdot l)]\mu+i_li_n(\frac{\delta\epsilon}{\epsilon}\epsilon)\\
    &=[\delta l\cdot n+\frac{\delta\epsilon}{\epsilon}]\mu.
\end{aligned}
\end{equation}
\
It is also useful to work out the variation of the expansion $\theta_+ = \frac{i_nd\mu}{\mu}$:
\begin{equation}\label{vartheta}
\begin{aligned}
    \delta\theta_+ = &\frac{\delta(i_n d\mu)}{\mu}-\frac{i_n d\mu}{\mu}\frac{\delta\mu}{\mu}\\
             &=\frac{i_n d\delta\mu}{\mu}-\theta_+\frac{\delta\mu}{\mu}\\
             &=i_n d(\frac{\delta\mu}{\mu}\mu)\mu^{-1}-\theta_+ \frac{\delta \mu}{\mu}\\
             &=\nabla_n(\frac{\delta\mu}{\mu})+\cancel{\frac{\delta\mu}{\mu}\theta_+}-\cancel{\theta_+ \frac{\delta \mu}{\mu}}            
\end{aligned}
\end{equation}
where these manipulations can be justified explicitly by writing $\mu=f\eta$, where $\eta$ is some fixed $D-2$ form annihilating $n$. Applying equations \eqref{dmu} and \eqref{vartheta} to equation \eqref{third term},
\begin{equation}
    \begin{aligned}
    -\nabla_n(\delta g^{cd}g_{cd})&=2\nabla_n(\frac{\delta\mu}{\mu}-\delta l\cdot n)\\
    &=2\nabla_n(\frac{\delta\mu}{\mu})-2\cancel{\nabla_n \delta l \cdot n}\\
    &=2\delta\theta_+.
    \end{aligned}
\end{equation}
Putting everything together into the master equation \eqref{charge}:
\begin{equation}\label{translation charge}
    \begin{aligned}
        -I_{\hat{\xi}} \Omega &=\int_{\dS} -i_\xi\theta_R^*\\
                      &=\int_{\dS} -\gamma_n[\cancel{\nabla_c(n_a\delta g^{ac})}-\nabla_a n_c \delta g^{ac}-\nabla_n (\delta g^{cd}g_{cd})]\mu\\
                      &=\int_{\dS} -\gamma_n[2\theta_+\frac{\delta\mu}{\mu}+2\delta\theta_+]\mu\\
                      &=\int_{\dS} -2\gamma_n\delta(\theta_+\mu)             
    \end{aligned}
\end{equation}
which is an integrable variation, yielding an observable that generates normal translations along $n$. The same arguments \textit{mutatis mutandis} in the $l$-adapted system produce a charge generating normal translations along $l$. This charge replaces $\theta_+$ with $\theta_-$, and has the opposite total sign (replace $n^b$ with $l^b$ in the first line of \eqref{insert}), because $l$ is oriented in the opposite direction relative to $\Sigma$.
\par
This expression was derived on a gauge fixed constraint surface of prephase space; however, the gauge invariant extension of $\eqref{translation charge}$ to all of prephase space retains the same form as long as the expansion is associated with the scaling frame discussed in section \ref{subsec:diffeo}. This is because that scaling frame transforms \textit{covariantly} under shearing and scaling transformations, which implies that the expansion is invariant under these transformations and thus that \eqref{translation charge} lifts to a gauge invariant expression.
\section{Charge Algebra}\label{sec:algebra}
We have now found a complete family of charges parametrizing the reduced phase space $\tilde{\mathcal{P}}$ for the diamond in 3 dimensional pure gravity\footnote{In fact, as we will see in the next section, it is slightly overcomplete due to the presence of Casimirs.}. These are summarized in table \ref{tab:charges} below. It remains to compute the Poisson brackets between them. These must represent (up to central extension) the Lie bracket between the flows they generate on $\tilde{\mathcal{P}}$, which is equivalent to representing the Lie bracket on $\mathcal{P}$ (with any consistent gauge fixing). It is known \cite{BT_soln_bracket,Riello_bracket,framework,einstein} that accounting for the solution dependence of the diffeomorphisms, this bracket is:
\begin{equation}\label{bracket}
    [\hat{\eta},\hat{\zeta}]_{\mathcal{P(\mathcal{R})}}=
   \widehat{ \llbracket \zeta,\eta \rrbracket}:=
    \widehat{[\zeta,\eta]}+I_{\hat{\eta}}\delta\zeta-I_{\hat{\zeta}}\delta\eta
\end{equation}
The representation condition reads:
\begin{equation}\label{representation}
    \{Q_\eta,Q_\zeta\}=Q_{\llbracket\eta,\zeta\rrbracket}+C_{\eta,\zeta}
\end{equation}
with $C$ a c-number constant on phase space. 
\par
In either of the extended gauge fixings we constructed in section \ref{subsec:norm}, the  vector fields $n$ and $l$ are fixed up to first order off the corner. This fixes the generating vectors of \textit{all} symmetries as parametrized in section \ref{sec:sym} up to first order. The boost and superrotation vector fields are already determined by just the values of $n$ and $l$ on the corner, and further fixing the first derivatives manifestly determines the normal translation generators. This means that the last two terms in equation \eqref{bracket} vanish up to first order off the corner, so they can be ignored in equation \eqref{representation}. The solution space brackets reduce to the spacetime Lie brackets of the generators (in the opposite order), which can be computed from the definitions\footnote{Technically this requires extending the defined generators to at least second order off the corner. Such extensions differ only by gauge transformations, so will not affect the result of the charge representation theorem.}. Alternatively, they can be easily deduced from the Poisson algebra via \eqref{representation}. The results are recorded in table \ref{tab:algebra}.\par

\begin{table}[]
    \centering
    
    \begin{tabular}{|c|c|}
    \hline
        Symmetry transformation & Charges \\
        \hline
         Boost & $\mathcal{Q_B}(\lambda)=\int_{\dS}-2\lambda\mu$\\
         \hline
         Superrotation & $\mathcal{Q_S}(\xi)=\int_{\dS}2\bar{\omega}(\xi)\mu$\\
         \hline
         Normal translations & $\mathcal{Q_+}(\gamma_n)=\int_{\dS}-2\gamma_n\theta_+\mu$\\
         & $\mathcal{Q_-}(\gamma_l)=\int_{\dS}2\gamma_l\theta_-\mu$\\
         \hline
    \end{tabular}
    \caption{Symmetry generators for three dimensional diamonds}
    \label{tab:charges}
\end{table}

Now we must compute the actual Poisson brackets. For each pair of charges we need only compute the bracket in one order, which can be chosen for convenience so that the variation of the first charge under the flow generated by the second is easy to find. Note that each charge is the integral of a top form, or alternatively a local density of weight $1$ on the corner against a fixed parameter. One might think that the variation can be computed by simply computing the Lie flow of this density, but we need to be careful; the charge densities are constructed out of geometric quantities such as the \Ha\;1-form and expansions which depend on the vectors $n$ and $l$ at the corner (but not on the extension away from it). In the original gauge fixing introduced in \ref{subsec:diffeo}, these are fixed structures that do not transform when the metric is flowed. Therefore when computing the flow it is necessary to hold $n$ and $l$ constant. Fortunately, however, with the exception of the boosts all of the symmetry vectors have vanishing Lie bracket with $n$ and $l$, so the Lie flow \textit{already} holds $l$ and $n$ constant. As long as we arrange for the boost charges to come first in any Poisson bracket, this means that it suffices to compute the Lie flow of the various densities.
\par
From here it is straightforward to compute most of the brackets. The action of any flow on the boost charges follows from the Lie flow of $\mu$, which is found immediately from the definition of the symmetries. The action of a superrotation flow on any charge is the integral of the Lie derivative of the relevant density with respect to the superrotation vector against the fixed parameter. Since the vector is tangent to the corner, this can be integrated by parts to move the Lie derivative onto the parameter instead. The remaining brackets are those between two normal translation charges, which demand more thought. These require computing the Lie derivative of the expansions under normal translations. Since they are scalars, we need only compute their derivatives at the corner in the corresponding directions, which are uniquely determined by the extension of the scaling frame along the horizons described in section \ref{sec:frames} --- however, for convenience we will use the stronger properties of the $n$-adapted gauge fixing of section \ref{subsec:norm}.  We will see that the derivatives follow from imposing the Einstein equation in the form \eqref{Riemann} at the corner, which is related to the arguments of \cite{anomaly,framework,einstein} that the equations of motion at the corner correspond to the closure of a charge algebra.\footnote{The authors of \cite{anomaly,framework,open_laurent,einstein} define generalized charges for diffeomorphisms and compute a notion of bracket between them, called the Barnich Trossaert bracket. They argue that the equations of motion at the corner --- in this case the Einstein equation --- cause this bracket to represent (up to a central extension given by a general formula) the solution space bracket. In general these charges will not generate the diffeomorphisms on phase space. When they do, the Barnich-Trossaert bracket between two such charges reduces to the Poisson bracket. Matching definitions, these correspond to the Noether charges in the present paper. For the boosts and superrotations, they are equal to the generators, but vanish for the normal translations. The central extension terms they find vanish identically with our choice of symplectic potential, so we expect that the Poisson brackets between boost and superrotation charges will agree exactly with the solution space Lie bracket. In fact this turns out to be true for all charges. \label{foot:BT}}
\par

To compute $\nabla_n\theta$, consider $R(n,\chi)n\cdot\chi$, where $\chi$ is a unit vector tangent to the corner extended away from it to first order by parallel transport along $n$ and $l$, so thar it remains a unit vector orthogonal to both. This contraction can be computed in two ways. First, explicitly:

\begin{equation}\label{Raychaudhuri}
    \begin{aligned}
        R(n,\chi)n\cdot\chi&=[\nabla_n\nabla_\chi n -\nabla_\chi\cancel{\nabla_n n}-\nabla_{[n,\chi]}n]\cdot \chi\\
        &=\nabla_n(\nabla_\chi n \cdot \chi)-\nabla_\chi n\cdot \cancel{\nabla_n\chi}-\nabla_{[\cancel{\nabla_n\chi}-\nabla_\chi n]}n\cdot \chi\\
        &=\nabla_n(K_+\chi\cdot\chi+\bar{\omega}(\chi)\cancel{n\cdot\chi})+\nabla_{[K_+\chi+\bar{\omega}(\chi)n]}n\cdot \chi\\
        &=\nabla_n(\theta_+\chi\cdot\chi)+ \theta_+\nabla_\chi n\cdot\chi+\bar{\omega}(\chi)\cancel{\nabla_n n}\cdot\chi\\
        &= \nabla_n\theta_++\theta_+^2.
    \end{aligned}
\end{equation}
 Secondly, it can be shown to vanish by equation \eqref{Riemann}. Therefore
\begin{equation}\label{theta+}
    \nabla_n\theta_+ = -\theta_+^2
\end{equation}
which is nothing but a special case of the familiar Raychaudhuri equation. In general dimension there will also be terms dependent on the shear. 
\par
Similarly, we can compute 
\begin{equation}\label{mixed}
    \begin{aligned}
        R(n,\chi)l\cdot\chi&=[\nabla_n\nabla_\chi l -\nabla_\chi\cancel{\nabla_n l}-\nabla_{[n,\chi]}l]\cdot \chi\\
        &=\nabla_n(\nabla_\chi l \cdot \chi)-\nabla_\chi l\cdot \cancel{\nabla_n\chi}-\nabla_{[\cancel{\nabla_n\chi}-\nabla_\chi n]}l\cdot \chi\\
        &=\nabla_n(K_{-}\chi\cdot\chi-\bar{\omega}(\chi)\cancel{l\cdot\chi})+\nabla_{[K_+\chi+\bar{\omega}(\chi)n]}l\cdot \chi\\
        &=\nabla_n(\theta_-\chi\cdot\chi)+ \theta_+\nabla_\chi l\cdot\chi+\bar{\omega}(\chi)\cancel{\nabla_n l}\cdot\chi\\
        &= \nabla_n\theta_-+\theta_+\theta_-,
    \end{aligned}
\end{equation}
as well as 
\begin{equation}\label{conformal flat}
    R(n,\chi)l\cdot \chi = \Lambda
\end{equation}
from \eqref {Riemann}. Altogether
\begin{equation}\label{theta-}
    \nabla_n\theta_-=-\theta_+\theta_--\Lambda,
\end{equation}
which is a special case of the cross-focusing equation introduced by Hayward in \cite{hayward_dynamics}. Equations \eqref{theta+} and \eqref{theta-} together with analogous relations for derivatives along $l$ are sufficient to determine the desired Poisson brackets\footnote{Equations \eqref{theta+} and \eqref{theta-} are also simple enough to solve explicitly in a neighbourhood of the corner. However the expansions may eventually diverge, if continued for long affine times. This corresponds to the occurrence of caustics\cite{Reall_caustics}. $\mathcal{R}$ was assumed to be a finite domain where this does not occur, in which \Np\;and \Nm\;are smooth.\label{foot:solving}}. The results are recorded in table \ref{tab:algebra}. They constitute an algebra, with solution independent structure constants. Note the explicit appearance of the cosmological constant in the last line, which gives the bracket of translation charges for distinct horizons. There are no central extension terms as in \eqref{representation}, so the Poisson algebra represents the Lie bracket algebra of the flows exactly. This arises from the fact that the charges are given by integrals of densities on the corner, which must transform under diffeomorphisms in a particular way. Since the extension terms $C_{\eta,\zeta}$ in \eqref{representation} are c-numbers, they would Poisson commute with the diffeomorphism generators and fail to transform appropriately. Thus they must vanish.

\begin{table}[]
    \centering
    \begin{tabular}{|c|c|c|}
    \hline
    Charge Pair & Solution Space Bracket & Poisson Bracket  \\
    \hline
       $\mathcal{Q_B}(\lambda_1),\mathcal{Q_B}(\lambda_2)$  &$\llbracket(\lambda_1,0,0,0),(\lambda_2,0,0,0)\rrbracket=0$ &$\{\mathcal{Q_B}(\lambda_1),\mathcal{Q_B}(\lambda_2)\}=0
       $\\
       \hline
       $\mathcal{Q_B}(\lambda),\mathcal{Q_S}(\xi)$  &$\llbracket(\lambda,0,0,0),(0,\xi,0,0)\rrbracket=(-\nabla_\xi\lambda,0,0,0)$ &$\{\mathcal{Q_B}(\lambda),\mathcal{Q_S}(\xi)\}=\mathcal{Q_B(-\nabla_\xi\lambda)}$\\
       \hline
       $\mathcal{Q_B}(\lambda),\mathcal{Q_\pm}(\gamma)$&$\llbracket(\lambda,0,0,0),(0,0,\gamma,0)\rrbracket=(0,0,\gamma\lambda,0)$ &$\{\mathcal{Q_B}(\lambda),\mathcal{Q_\pm}(\gamma)\}=\pm\mathcal{Q_\pm(\gamma\lambda)}$\\
       \hline
       $\mathcal{Q_S}(\xi_1),\mathcal{Q_S}(\xi_2)$ & 
       $\llbracket(0,\xi_1,0,0),(0,\xi_2,0,0)\rrbracket=(0,[\xi_1,\xi_2],0,0)$&
       $\{\mathcal{Q_S}(\xi_1),\mathcal{Q_S}(\xi_2)\}=\mathcal{Q_S}([\xi_1,\xi_2])$\\
       \hline
       $\mathcal{Q}_\pm(\gamma),\mathcal{Q_S}(\xi)$&
       $\llbracket(0,0,\gamma,0),(0,0,\xi,0)]=(0,0,-\nabla_\xi\gamma,0)$&
       $\{\mathcal{Q}_\pm(\gamma),\mathcal{Q_S}(\xi)\}=\mathcal{Q}_\pm(-\nabla_\xi \gamma)$\\
       \hline
       $\mathcal{Q}_\pm(\gamma_1),\mathcal{Q}_\pm(\gamma_2)$&$\llbracket(0,0,\gamma_1,0),(0,0,\gamma_2,0)\rrbracket=0$&$\{\mathcal{Q}_\pm(\gamma_1),\mathcal{Q}_\pm(\gamma_2)\}=0$\\
       \hline
     $\mathcal{Q}_\pm(\gamma_1),\mathcal{Q}_\mp(\gamma_2)$&$\llbracket(0,0,\gamma_1,0),(0,0,0,\gamma_2)\rrbracket=(-\gamma_1\gamma_2\Lambda,0,0,0)$&$\{\mathcal{Q}_\pm(\gamma_1),\mathcal{Q}_\mp(\gamma_2)\}=\mathcal{Q_B}(\mp \gamma_1 \gamma_2 \Lambda)$\\
    \hline
    \end{tabular}
    \caption{Charge brackets of three dimensional diamonds. Column 2 shows the solution space brackets of the flows, described by the 4-tuple $(\lambda,\xi,\gamma_n,\gamma_l)$ of parameters characterizing a given symmetry --- for simplicity only one of the flows for $\mathcal{Q}_+,\mathcal{Q}_-$ is shown when they are easy to interchange. Column 3 shows the Poisson brackets of the corresponding charges.}
    \label{tab:algebra}
\end{table}
\par
The charge algebra has a simple structure --- it is $\mathfrak{diff}(\mathcal{S})\;\tilde{\oplus}\;\mathfrak{sl}(2,\mathbb{R})^\mathcal{S}$, where $\tilde{\oplus}$ denotes the semidirect sum of Lie algebras \cite{semidirect} (this notation is not standard). The Lie ideal $\mathfrak{sl}(2,\mathbb{R})^\mathcal{S}$ corresponds to the subalgebra of boosts and normal translations, interpreted as a vector space of continuous maps from $\mathcal{S}$ to $\mathfrak{sl}(2,\mathbb{R})$. This follows because the algebra of boosts and translations "at a point" is isomorphic to $\mathfrak{sl}(2,\mathbb{R})$\footnote{Thanks to Antony Speranza for pointing this out. In the flat space limit this is no longer true --- instead it is isomorphic the Poincare algebra in two spacetime dimensions. This case will be treated implicitly as the limit as $\Lambda\rightarrow 0$.}. The $\mathfrak{diff}(\mathcal{S})$ factor of course refers to the superrotations, and acts on the ideal in the obvious way. This algebra is isomorphic to the corner symmetry algebra found in \cite{subsystem}, but it does \textit{not} have the same interpretation --- the flows generated by the $\mathfrak{sl}(2,\mathbb{R})$ algebras in \cite{subsystem} correspond to boosts and transformations that tilt the horizons. Since we have fixed that the horizons are null, the latter do not exist in our algebra. Their algebra also does not include normal translations at all. It is tempting to follow those authors and conclude that we have a symmetry group isomorphic to $\mathrm{Diff}^+(\mathcal{S})\ltimes SL(2,\mathbb{R})^{\mathcal{S}}$, but this is not possible because the normal translation flows, for large flow parameter, will violate the smoothness assumptions on $\mathcal{R}$. The problem is that translating along \Np\;or \Nm\;beyond $\mathcal{R}$ can lead to crossings; see footnote \ref{foot:solving}.
\par
The charge algebra itself is what we care about. A basis of the algebra constitutes a possibly overcomplete set of coordinates of the reduced phase space. The interesting aspect is that they have a dual geometric interpretation; the \textit{values} of the charges are features of the geometry at the corner, and the \textit{flows} they generate are diffeomorphisms. By flowing with the translation charges, these geometric quantities can be computed for another cut along \Np\;or\;\Nm. From there it is possible to translate into the interior of the diamond. This allows one to construct the geometry of the diamond, at least inside $\mathcal{R}$, purely by knowing the values of the charges at the corner! Note that it is also possible to translate \textit{backwards} along \Np\; or \Nm\;by using negative values of the parameters $\gamma_n$ or $\gamma_l$. This constructs an extension of the spacetime \textit{beyond} the diamond, even though that was not part of the original setup. This is a result of the fact that there are no propagating degrees of freedom in 2+1 dimensions, so the geometry of the corner really determines the geometry in it vicinity. In a certain sense, this means that the system realizes the \textit{holographic principle}; all data is encoded on a codimension-2, rather than codimension-1, surface \cite{harlow_tasi}.
\par
Some remarks on the generalization of this picture are in order. In arbitrary dimensions the superrotation and boost charges and their brackets are unchanged. However the normal translations no longer give rise to integrable charges. Examining equation \eqref{insert} and the subsequent simplification, it is clear that the only correction arises from the nonvanishing of the shear, resulting in an extra term $\sigma_{ac}\delta q^{ac}\mu$ in the insertion of the generator into the symplectic form. Since the shear is a projected tensor intrinsic to the corner, it (along with the full corner metric) provide additional observables defined with respect to the corner coordinates. The nonintegrable charge variation for the normal translation can be expressed in terms of the variations of the expansion, shear, and metric on the corner, so a similar geometric parametrization of the "edge mode" data is available. However, the brackets among these observables no longer close into an algebra because the equations of motion at the corner involve additional information; for example, equation \eqref{conformal flat} no longer applies because the Riemann tensor does not take the form \eqref{Riemann}, as the metric is not guaranteed to be conformally flat. This corresponds to the physical expectation that graviational waves from the diamond interior can propagate to the horizon, so data on the corner should not evolve as an isolated system. The same holds true if there are nontrivial matter fields, since their excitations can leave the diamond. 
\section{Casimirs and Coadjoint Orbits}\label{sec:coadjoint}
As has been noted, the algebra in table \ref{tab:algebra} may be an overcomplete set of coordinates on the reduced phase space. This will happen if on $\tilde{\mathcal{P}}$ the charges satisfy further relations among themselves, so that not all possible values are realized. Indeed this is the case. To find these relations, we will follow Kirillov's coadjoint orbit method \cite{Kirillov1,Kirillov2,Kirillov3}, which has been previously studied for the original corner symmetry algebra of \cite{subsystem} in \cite{hydrodynamics}. Note that while that algebra is isomorphic to ours in three dimensions, \cite{hydrodynamics} carries out the orbit analysis in four dimensions. Because of this, although structure of the logic will largely be the same, the results turn out to be quite different.
\par
The main idea of the orbit method is to consider how the phase space is mapped into the vector space $\mathcal{A}$ of all possible values that can be assigned simultaneously and linearly to the charges in table $\ref{tab:charges}$, without regard to whether these correspond to valid solutions in the phase space. $\mathcal{A}$ is called the \textit{coadjoint representation} of the symmetry algebra, because any point in it can be naturally interpreted as an element of the dual space to the algebra; given an element of the algebra, labeled by the parameter tuple $(\lambda, \xi, \gamma_n, \gamma_l)$, the point produces a real number by evaluating the associated charge. Since the coadjoint representation is not restricted to values corresponding to an actual solution, the action of the algebra in table \ref{tab:algebra} generates\footnote{This is not the same as \textit{exponentiates}, due to infinite dimensional subtleties. The exponential map of the Lie algebra $\mathfrak{diff}(\mathcal{S})$ is not locally surjective onto a neighbourhood of the identity in $\mathrm{Diff}^+(\mathcal{S})$ \cite{Milnor_infinite}. The inclusion of time dependent parameters is crucial in order to obtain the full group.} a left group action on $\mathcal{A}$, which can be computed by letting the parameter tuple depend on a path time and integrating the infinitesimal transformation. This group is not necessarily $\mathrm{Diff}^+(\mathcal{S})\ltimes SL(2,\mathbb{R})^{\mathcal{S}}$, because a path that might appear to correspond to a noncontractible loop in that group could turn out to have nontrivial action. The most general group that can be generated by integrating the algebra is the universal cover $G=\mathrm{Diff}^+(\mathcal{S})\ltimes \widetilde{SL}(2,\mathbb{R})^\mathcal{S}$, where the normal subgroup is the appropriately defined tensor power of the universal cover of $SL(2,\mathbb{R})$; any other such group is covered by a homomorphism from $G$, which implies that the action can always be lifted to $G$. We will take this to be the symmetry group.

\par The left $G$-action can be written explicitly in terms of the transformation of the local charge densities. Define the covector density
$\mathcal{Q_{\bar{\omega}}}=2\bar{\omega}\mu$ and the tuple of generator densities $\mathcal{Q}_a$ with $a\in \{\mathcal{B},+,-\}$, which can be interpreted as a densitized dual to $\mathfrak{sl}(2,\mathbb{R})$. Under a pure integrated superrotation,  denoted by a map $\varphi\in \mathrm{Diff}^+(\mathcal{S})$, these transform by pullback:
\begin{equation}\label{diffeo_action}
L_\varphi(\qw)= \varphi^*(\qw) \qquad L_\varphi(\qa)= \varphi^* (\qa).
\end {equation}
The action of a pure $\widetilde{SL}(2,\mathbb{R})$ transformation, denoted by an element $g(s)\in \widetilde{SL}(2,\mathbb{R})^\mathcal{S}$. is

\begin{equation}
L_g(\qw) = \qw+(g^{-1} dg)^a \qa  \qquad L_g(\qa) = (\mathrm{Ad}_{g^{-1}})^*\qa.
\end{equation}
The second term in the first equation expresses the contraction of the $\mathfrak{sl}(2,\mathbb{R})$-valued covector $g^{-1}dg$ with the densitized dual $\qa$, and the second equation expresses the pullback of $\qa$ under the adjoint action of $g^{-1}$ on $\mathfrak{sl}(2,\mathbb{R})$ --- the inverse is because a generator acts from the right in Poisson brackets, so the action induced by the $\mathfrak{sl}(2,\mathbb{R})$ algebra is the opposite of the usual adjoint action.
\par
$\mathcal{A}$ is foliated by submanifolds on which $G$ acts transitively, called \textit{coadjoint orbits}. A fundamental observation is that $\mathcal{A}$ is a Poisson manifold, and the orbits carry a natural symplectic structure compatible with the Poisson structure \cite{Kirillov3,Kostant1,Kostant2,Souriau}. Evaluating the charges at any point of the reduced phase space maps it into the coadoint representation, and since we know that diffeomorphisms act transitively on phase space (see section \ref{sec:cov}) $\tilde{\mathcal{P}}$ must be mapped into a single orbit.
\par
Coadjoint orbits are in general labeled by \textit{Casimir invariants}, functions of the charges that Poisson commute with everything, and possibly also discrete variables (referred to as generalized Casimirs in \cite{rodrigo_quantum}). On any given coadjoint orbit, all of these must be constant, which enforces some relations among the charges on that orbit. For the algebra under consideration, there are exactly two Casimirs and one discrete label. To find them, first consider the formal $\mathfrak{sl}(2,\mathbb{R})$ algebra \textit{at a point}, which is 
\begin{equation}\label{sl2r}
    \mathcal{\{Q_B,Q_+\}=Q_+\;\;\;\;\;\;\{Q_B,Q_-\}=-Q_-}\;\;\;\;\;\;\mathcal{\{Q_+,Q_-\}}=-\Lambda\mathcal{Q_B}
\end{equation}    
which can be integrated against smearing functions to obtain a proper Poisson algebra. This has a Casimir invariant
\begin{equation}\label{sl2r_casimir}
    \mathcal{C}_{\widetilde{SL}(2,\mathbb{R})}:=\Lambda Q_B^2-2Q_+Q_-=4\mu^2(\Lambda+2\theta_+\theta_-)
\end{equation}
that is a density of weight $2$. This density is said to be positive if it yields a positive number when evaluated on a positively oriented tangent vector to the corner. For the orbit analysis, we will restrict to the region of $\mathcal{A}$ where the Casimir density is positive at all points of the corner, which is closed under the action of $G$. It is easy to see that $\tilde{\mathcal{P}}$ maps into this region, since for small loops embedded in an MSS the expansions are both negative and large, so the second term in \eqref{sl2r_casimir} is positive and dominates the first. 
\par
Positivity of the density allows us to take the square root and obtain another Casimir density of weight $1$:
\begin{equation}\label{alpha}
    \alpha:=\sqrt{\mathcal{C}_{\widetilde{SL}(2,\mathbb{R})}}=2|\mu|\sqrt{\Lambda+2\theta_+\theta_-}
\end{equation}
where $|\mu|$ is just $\mu$ with a sign to make it positive as necessary. The integral
\begin{equation}\label{C_1}
    \mathcal{C}_1=\int_\mathcal{S} \alpha
\end{equation}
is manifestly invariant under both superrotations and $\widetilde{SL}(2,\mathbb{R})$ transformations, and is therefore a Casimir of the total algebra, which corresponds to the "total mass Casimir" of \cite{hydrodynamics}. Note that this quantity is similar, but not equal to, the Hawking energy \cite{Hawking_energy} of $\mathcal{S}$.
\par
Now following \cite{hydrodynamics}, we note that it is possible to transform $\qa$ to satisfy
\begin{equation}\label{reduction_1}
    \mathcal{Q_B}=0 \qquad \mathcal{Q_+}=-\mathcal{Q_-}=\pm\sqrt{2}\alpha
\end{equation}
 by acting with a pure $\widetilde{SL}(2,\mathbb{R})$ transformation; the sign is a discrete label depending on the orbit. 
 On solution space, this corresponds to applying a boost so that the expansions become equal and using normal translations to contract the loop to a "point". However, the argument does not rely on the geometry of the original solutions, only on algebraic properties of $\widetilde{SL}(2,\mathbb{R})$. There are many transformations that achieve \eqref{reduction_1}; any two are connected by a transformation obtained by exponentiating a generator that is a smearing of $\mathcal{Q}_+-\mathcal{Q}_-$.
\par
The second Casimir invariant can be defined in terms of any $g(s)$ guaranteeing \eqref{reduction_1}, and then shown to be independent of the particular choice. For a point in $\mathcal{A}$, consider the unique superrotation vector $\xi_\alpha$ such that $\alpha(\xi_\alpha)=1$ everywhere on the corner. The Casimir is given by 

\begin{equation}\label{C_2}
\begin{aligned}
    \mathcal{C}_2&=\int_\mathcal{S} L_g(\qw)(\xi_\alpha) \\
                 &=\int_\mathcal{S} [\qw+(g^{-1}dg)^a \qa ](\xi_\alpha).
 \end{aligned}   
\end{equation}
Since $\qw$ is a covector density of weight 1 and therefore equivalent to a symmetric $(0,2)$-tensor on $\mathcal{S}$, it does not matter which index is contracted with $\xi_\alpha$. This quantity replaces a hierarchy of Casimir invariants constructed out of a "dressed vorticity" in \cite{hydrodynamics}, as these all vanish when $D=3$. Note the apparent relation of $\mathcal{C}_2$ to a superrotation by $\xi_\alpha$. In fact, the Poisson bracket of any charge with $\mathcal{C}_2$ at a point in $\mathcal{A}$ is equal to the bracket with $\mathcal{Q_S}(\xi_\alpha)$, plus the bracket with a $\widetilde{SL}(2,\mathbb{R})$ generator. Once it is established that $\mathcal{C}_2$ is a Casimir, it is apparent that the effect of this $\widetilde{SL}(2,\mathbb{R})$ generator is to cancel that of $\mathcal{Q_S}(\xi_\alpha)$.
\par
To see that this is independent of the choice of $g$, consider perturbing $g$ to $g'(\epsilon)g$, where $g'$ preserves \eqref{reduction_1}. Then to first order in $\epsilon$, $g'=1+\epsilon \chi g$ for some $\chi\in\mathfrak{sl}(2,\mathbb{R})^\mathcal{S}$ corresponding to a smearing of $\mathcal{Q}_{+}-\mathcal{Q}_{-}$ by a scalar function $\gamma$. The derivative of $L_{g'g}(\qw)$ is
\begin{equation}\label{perturbation}
    \begin{aligned}
        \delta_{\epsilon}L_{g'g}(\qw)&= \delta_\epsilon L_{g'}(L_g(\qw)) \\
                                      &= \delta_\epsilon[L_g(\qw)+(g'^{-1}dg')^a L_g(\qa)]\\
                                      &= d\chi^{a}L_g(\qa)\\
                                      &= d\gamma(\pm 2\sqrt{2}\alpha)
    \end{aligned}
\end{equation}
where on the last line we re-express $\chi$  in terms of $\gamma$ and carry out the contraction using equation \eqref{reduction_1}. Upon inserting $\xi_\alpha$ into the second index and integrating, we find that $\delta_\epsilon \mathcal{C}_2=0$, proving that the choice of $g$ does not matter and $\mathcal{C}_2$ has a unique value at each point of $\mathcal{A}$. This also implies, via the composition law for the group action, that $\mathcal{C}_2$ is invariant under $\widetilde{SL}(2,\mathbb{R})$ transformations. Furthermore, since all quantities in the integrand of \eqref{C_2} --- $g$, $\qw$, and $\xi_\alpha$ --- transform by pullback under a superrotation, the integral is preserved under superrotations as well. This completes the argument that $\mathcal{C}_2$ is a Casimir invariant. 

\par
Now we will show that an orbit is completely determined by the values of the Casimirs $\mathcal{C}_1$ and $\mathcal{C}_2$, as well as the sign in equation \eqref{reduction_1}. Given a point in $\mathcal{A}$, we will apply a sequence of transformations to convert it into an orbit representative which is purely determined by the stated invariants. First, fix a positive reference density $\tilde{\alpha}$ with total integral equal to $\mathcal{C}_1$ and apply a superrotation to set $\alpha=\tilde{\alpha}$ . Then apply a $\widetilde{SL}(2,\mathbb{R})$ transformation to guarantee that \eqref{reduction_1} holds. This fixes $\qa$ in terms of $\tilde{\alpha}$ and the discrete sign, and sets 
\begin{equation}\label{reduction_2}
\int_\mathcal{S} \qw(\xi_{\tilde{\alpha}})=\mathcal{C}_2.
\end{equation}
This transformation can be varied as in equation \eqref{perturbation}. Writing $\qw=\beta\tilde{\alpha}$, where $\beta$ is a covector, this variation induces the transformation
\begin{equation}
    \beta\rightarrow \beta+d\gamma
\end{equation}
for an arbitrary choice of scalar function $\gamma$. This allows $\beta$ to be set to any covector on $\mathcal{S}$ that integrates to $\mathcal{C}_2$; we can set
\begin{equation}\label{reduction_3}
\beta = \frac{\mathcal{C}_2}{\mathcal{C}_1}\tilde{\alpha},
\end{equation}
which concludes the process of fixing $\qw$ and $\qa$ in terms of the invariants. Hence $\mathcal{C}_1$, $\mathcal{C}_2$, and the $\pm$ of equation \eqref{reduction_1} together completely determine a coadjoint orbit.\footnote{The completeness of this set of invariants has an interesting consequence. Any diamond admits a 6-dimensional family of infinitesimal isometries, which may be thought of as inherited from an MSS upon embedding. Each isometry vector field has associated coefficients $(\lambda,\xi,\gamma_n,\gamma_l)$ that can be read off from its behaviour up to first order off the corner. The charge variation obtained by inserting the isometric flow into the presymplectic form must, on the one hand, be equal to the linear combination of the charge variations computed in section \ref{sec:charge}, according to those coefficients; on the other hand, it must be zero. This implies that those coefficients correspond to a linear combination of $\delta \mathcal{C}_1$ and $\delta \mathcal{C}_2$.}
\par
To find the particular orbit that the phase space is mapped into, we can compute the invariants by choosing a simple solution. Consider a static timeslicing of an MSS, and a diamond that is the domain of dependence of a rotationally symmetric disk on one slice. Also let the scaling frame be rotationally symmetric. The corner loop can be contracted to a point while maintaining symmetry by applying normal translations, which implements \eqref{reduction_1}. Although $\theta_{+,-}$ diverge and $\mu$ vanishes in this limit, the limit of the relevant products is finite and straightforward to compute. One finds that $\mathcal{C}_1=4\pi$, and that the discrete sign is $+$. By rotational symmetry, $\bar{\omega}$ vanishes for any such diamond. Since $\mathcal{C}_2$ is the integral of $\qw$ in the limit as the loop is shrunk, it also vanishes. 
\par
Phase space will be mapped onto some region of the orbit determined by these values of the invariants. This region will be connected, since the action of symmetries on phase space is transitive. Two distinct points of the reduced phase space could be mapped to the same point on the orbit; however, this requires that the path connecting the two points map to a noncontractible loop on the orbit. If the orbit is simply connected, this cannot happen. To decide the issue one way or the other, one needs to compute the orbit's fundamental group. Since we will not pursue this problem here, we must content ourselves with a parametrization of the phase space up to a possible discrete identification of points.
\par
Do the other coadjoint orbits have a physical interpretation? The results of the orbit analysis suggest that they might. It seems natural to speculate that there is some way to extend the treatment to domains of dependence that contain topological nontrivialities, such as BTZ black holes \cite{BTZ}, or other sources, like point particles \cite{particles}. Since the computation based on contracting the diamond to a point no longer applies, one expects that these would correspond to different values of the invariants. In particular, the relation of $\mathcal{C}_1$ to the Hawking energy suggests that it somehow captures the "energy" of the black hole or particle, and the similarity of $\mathcal{C}_2$ to a rotation generator suggests that it detects "angular momentum", which is consistent with the fact that it vanishes for the original phase space. Pursuing such an extension is an attractive direction for future research.

\section{Discussion}\label{sec:discuss}
\subsection{Relation to Previous Work}\label{subsec:relation}
In summation, the main results of the present work are the parametrization of the reduced phase space of 2+1 dimensional gravity in terms of the charges listed in table \ref{tab:charges}, the computation of their Poisson algebra shown in table \ref{tab:algebra}, and the complete set of Casimir invariants found in \ref{sec:coadjoint}. The charges are both geometric observables on the corner and generators of geometric flows, which enables one to use their values in a given state to reconstruct the spacetime. They are defined with respect to a fixed set of corner coordinates and a scaling frame with a particular solution dependence. Changing the structures but retaining the solution dependence only reparametrizes the charges in an obvious way --- in particular, a change of corner coordinates is equivalent to a finite superrotation, and a change of scaling frame is equivalent to a finite boost. This endows the charges with a natural classification --- those for which $\xi=0$ and $\lambda,\gamma_n,\gamma_l$ are uniform on the corner are independent of the corner coordinates, and the pure boost charges are independent of the scaling frame. The \textit{uniform pure boost} has a special status, as it is independent of both. If one were to restore factors of $16\pi G$ and set $\lambda=2\pi$, one would find that it is equal to $-\frac{A}{4G}$, which is minus the Bekenstein-Hawking entropy. Of course, the Casimir invariants also have the property of being invariant under such reparametrizations, although they are not members of the charge algebra.
\par
In this section I will compare these results and methods to prior work. The foundational paper on corner symmetries at finite boundaries, \cite{subsystem}, and all of the subsequent literature differs from the present work in at least some respects, as far as I am aware --- some more than others. First, I'll consider methodological differences. In \cite{subsystem}, Donnelly and Freidel introduced the use of auxiliary embedding fields to isolate edge mode degrees of freedom, which has since been employed frequently. However, as pointed out in \cite{ambiguity}, these are redundant variables which can eliminated via passing to a "unitary gauge" which is completely equivalent to the setup in this paper. 
\par
A far more substantive difference lies in the fact that papers such as \cite{subsystem,local, isolated} focus on solution-independent diffeomorphisms, which are fixed with respect to the background manifold. Of course, this has since been extended, and there are many papers that embrace the solution dependence. In particular the general formalism described in \cite{anomaly,framework} does so, and those results are used in this paper. What is new, to my knowledge, is the organization of the solution dependence in terms of specific external structures, and the corresponding classification of the charges. Similar structures were used in \cite{fixed_area} to construct a version of a subregion phase space. Those were a conformal frame on the normal plane (which corresponds to the notion of scaling frame employed in this paper) and an induced metric on the corner, which is not fixed in this paper. However, the motivation of organizing the diffeomorphisms according to these structures is not present there. 
\par
Now turning to the results, the main distinctions from existing work pertain to the normal translation charges and the algebra that they form with the boosts. In general dimension, there are no integrable Hamiltonian charges which actually generate flows that change the location of the corner. However, in \cite{symm_null} integrable charges for these have been obtained via the Wald-Zoupas method \cite{wald_zoupas}, which focuses on conservation properties rather than the flows generated. These are specified to the case of causal diamonds in \cite{symm_causal}, and agree up to sign conventions with the expansion charges found in this paper. An extension of the charge bracket to include these, referred to as a generalized Barnich-Troessaert bracket in reference to \cite{BT}, has been defined and studied in \cite{anomaly, framework, open_laurent, einstein}. This represents the solution space Lie bracket of diffeomorphisms up to central extension (see footnote \ref{foot:BT}), However, this is not a Poisson bracket, and is thus incompatible with the conventional understanding of canonical quantization. 
\par
In \cite{Wieland} Wieland argued that imposing constraints on a gravitational phase space at null infinity to remove radiative degrees of freedom leaves a residual phase space consisting of only diffeomorphism edge modes, on which there are integrable translation charges and the aforementioned generalized BT-bracket reduces to a Poisson bracket. The present study essentially realizes this idea explicitly at finite boundaries in the case of 2+1-D gravity, where there are no radiative modes to begin with. Furthermore, the logic in the general case of Wieland is perhaps made more transparent; if all tangent directions to the phase space are associated to diffeomorphisms, there must be some charges that generate them. The problem is just to arrange the solution dependence in a convenient way to obtain a simple parametrization.
\par
There exist other studies of similar phase spaces for general relativity which are relevant. Some of these \cite{geiller1,geiller2,geiller3,geiller4} are studied in first-order variables incorporating a dynamical tetrad. I am not aware of a previous derivation of the general integrability of the normal translations (without boundary conditions much stronger than the ones assumed here), but the results here should generalize immediately. This is because the presymplectic form of general relativity in first order variables, as usually derived, differs from \eqref{symp} by a corner term that only has the effect of adding local Lorentz charges \cite{tetrad}. This corner term can be subtracted off or not, in which case these charges have to be computed and added to the existing charge algebra. It is interesting to note that the structures we've introduced at the corner can be viewed as a double-null tetrad with a specified solution dependence. 
\par
In a similar vein there is \cite{CLP}. The authors extend the phase space with the aforementioned embedding fields introduced in \cite{subsystem}. However, they alter the symplectic structure by adding a corner term, in a manner analogous to how the most obvious symplectic structure for GR in tetrad variables differs from that in terms of the metric. The effect is to introduce additional charges that generate diffemorphisms acting on both the dynamical and embedding fields. Such transformations were gauge in \cite{subsystem}, but are now not gauge for diffeomorphisms nonvanishing at the boundary. This implies that it is not possible to fix the embedding fields in a "unitary gauge", so they are no longer purely auxiliary variables \cite{ambiguity}. It is interesting to note that all of the new charges are integrable generators, including those corresponding to normal translations. However, they are similar to the Lorentz charges discussed above, in that they generate the transformation of new variables rather than actually changing the situation for the original diffeomorphisms acting on dynamical fields only, which are still nonintegrable. Therefore the interpretation is  not so close to our paper, which makes special use of the three dimensional behaviour to construct charges for the physical diffeomorphisms.
\par
There are two families of work which are more materially related to this paper. The first consists of \cite{slicings1,slicings2} in particular, with background material provided by \cite{slicings3,slicings4,slicings5}. What these authors refer to as "slicings" are essentially choices of solution dependence. The main idea of those papers is the same as this one --- in 2+1 dimensions, since all degrees of freedom are diffeomorphism edge modes, it is possible to construct a complete family of generators, including ones which move the corner. However, the geometric setup of is fundamentally different; they consider a single fixed null boundary and coordinates with specific properties in a neighbourhood of it, that can be matched onto the extended gauge fixing coordinates of section \ref{subsec:norm}. In terms of the metric expressed in these coordinates, they obtain 3 charge families which are functionals of parameters on the null boundary, not just the corner; in our notation, they are integrals of densities over \Nm. They constitute an overcomplete parametrization of the reduced phase space, and can be obtained from our corner charges using the equations of motion. The Poisson algebra between them is in fact \textit{independent} of the equations of motion (and thus the cosmological constant), which are treated as additional inputs. Comparatively, our charge algebra contains all the information of the theory and is represented on the codimension-two corner alone. As noted in section \ref{sec:algebra}, this means that it can be viewed as \textit{holographic}, whereas the charge algebra of those papers is not.
\par
Finally, there are the papers that motivated this project, \cite{2+1Diamond} and its more exhaustive sequels \cite{rodrigo_classical,rodrigo_quantum}. These also study 2+1 dimensional causal diamonds, but via noncovariant methods and only for $\Lambda\leq0$. The phase space is described in terms of ADM variables on a slice that is required to have a fixed mean extrinsic curvature and corner metric. These assumptions allow the constraints to be solved and gauge directions to be quotiented out, resulting in a reduced phase space which is the cotangent bundle of $\mathrm{Diff}^+(\mathcal{S})/PSL(2,\mathbb{R})$. This is then parametrized by two charge families which are the integrals of local densities, chosen by a coadjoint orbit method. These are labeled $P$ and $Q$, and form a centrally extended $\mathfrak{bms_3}$ algebra.\par

This phase space is actually a constraint surface of ours. Fixing the corner metric amounts to fixing the boost generators as a family of first class constraints. The resulting constrained space is parametrized by the remaining generators, but admits boost orbits which are symplectically degenerate. Gauge fixing the boosts amounts to imposing an additional family of constraints which, together with the boosts, are second class. There are many ways to do this --- for example, one could simply fix the expansions along one horizon. A more complicated route is to impose the CMC condition on a given slice at the level of the prephase space, which will descend to a family of boost-fixing constraints on reduced phase space (since boosts in general change the extrinsic curvature). The resulting phase space is exactly that found in \cite{2+1Diamond}. The $P$-charges are actually equal to the diffeomorphism charges we found, and the $Q$-charges are complicated functions that involve the expansions. However their brackets are \textit{not} given by reparametrizing those in table \ref{tab:algebra}, since the symplectic structure is deformed by the second class constraints; one has to compute the Dirac bracket. There may also be a deformation due to the use of the ADM symplectic structure, which differs from \eqref{symp} by a corner term (see \cite{mixed_boundary} for an explicit expression). Therefore their algebra does not have the dual geometric interpretation that is the key property of the one found in this paper, as can be seen from the central extension, which prevents the $P$-charges from acting as actual diffeomorphism generators on the $Q$-charges.\footnote{However, these charges have the property that their flows take points in the phase space to other points in the phase space for arbitrarily large flow times, meaning that they actually do define a symmetry group acting on the entire phase space. This was important in that work since this group formed the basis of the quantization procedure.}
\par
Our work has concentrated entirely on finite causal diamonds, and ignored the case of asymptotic AdS-like boundaries. As mentioned in the introduction, the symmetries at such boundaries and their implications for the phase space have long since been understood \cite{Brown_Henneaux,Maloney_Witten}. Recently this description has even been extended to nontrivial interior topologies \cite{universal_phase}, and to some finite timelike boundaries \cite{box,boundary_gravitons}. It would be interesting to understand the connection between those results and ours. Seemingly, this would involve taking the limit as the diamond corner approaches infinite size when $\Lambda<0$. Our algebra is rather larger than the original double Virasoro algebra found by Brown and Henneaux, and does not have any central extension. This probably arises from the fact that we have imposed very little in the way of boundary conditions, and therefore do not reproduce the asymptotic structure of AdS in the above limit. Phase spaces allowing for variable asymptotic structures were studied in \cite{leaky}, where it was found that these are "leaky" in that there is nonvanishing symplectic flux through the conformal boundaries. Our phase space is also leaky in this sense, and it may be that the asymptotic limit is a three dimensional case of such a phase space. If so, the original Brown-Henneaux algebra would arise as the Dirac algebra on a constraint surface, just as with the algebras found in \cite{2+1Diamond,rodrigo_classical,rodrigo_quantum}.

\subsection{Remarks on Quantization}\label{quant}
\par
Although the focus of this paper is on the classical theory, the results open an interesting possibility for quantum gravity. Since the charges are geometric features of the corner, if the algebra is quantized the resulting charges may be given a similar interpretation by analogy. Furthermore since the classical flows of the charges can be used to translate the corner into the interior (see section \ref{sec:algebra}), the exponentiated unitary action seems to allow the construction of a \textit{quantum spacetime}, on which the corner charges allow us to probe the geometry. This is a unique property of the newly constructed algebra, that was not possible for the constrained 
 approach of \cite{2+1Diamond,rodrigo_classical,rodrigo_quantum} or other approaches such as \cite{hydrodynamics,universal_corner,matrix_quantization,corner_quantum_geometry} which discuss quantum representations of the corner symmetry algebra, since these do not include Hamiltonian generators of the normal translations. To conclude this paper, I will give a brief overview of what such a quantization may look like. The details will be left to the future.
 \par
 A quantization will irreducibly represent the charges as self-adjoint operators on a Hilbert space, such that the commutators are determined by the Poisson brackets up to order $\hbar$:
 \begin{equation}\label{quantum}
     \mathcal{[\hat{Q}_\eta,\hat{Q}_\zeta]}=i\mathcal{\hbar\widehat{\{Q_\eta,Q_\zeta\}}}+O(\hbar^2).
 \end{equation}
 However, the $O(\hbar^2)$ terms must vanish if the algebra is to retain its geometric interpretation, for the same reason that central extensions were forbidden in section \ref{sec:algebra}; the generators are geometric quantities that must transform appropriately under diffeomorphisms. We will proceed under this premise, which implies that the algebra retains the structure of a semidirect sum. Since the algebra takes center stage, instead of a phase space, by the arguments of \ref{sec:coadjoint} it generates a unitary action of $G=\mathrm{Diff}^+(\mathcal{S})\ltimes \widetilde{SL}(2,\mathbb{R})^\mathcal{S}$ on the Hilbert space. Mackey's theory, which can be thought of as an quantum extension of the orbit method, provides a picture of what the representations look like, although the infinite dimensionality of the algebra means that it is only heuristic. This method is discussed in \cite{subsystem} and developed in more detail in \cite{hydrodynamics,universal_corner,matrix_quantization} for the original corner symmetry algebra, which is isomorphic to ours, although the difference in interpretation induces some subtle distinctions.
\par
Schematically, the approach is as follows: first, decompose the Hilbert space $\mathcal{H}$ into a direct sum of irreducible unitary representations of the normal subgroup $\tilde{SL}(2,\mathbb{R})^S$. Each such representation will be a tensor power of unitary irreps of $\tilde{SL}(2,\mathbb{R})$, labeled by $\alpha$, the square root of the (appropriately symmetrized for self-adjointness) $\mathfrak{sl}(2,\mathbb{R})$ Casimir density at each point. Because of the uncountable number of points on the corner, the actual operators are technically integrals of local densities and some kind of Fock space-like construction is required to get a separable Hilbert space; we will return to this point. For the moment we proceed with the picture of an infinite direct sum of infinite tensor products:
\begin{equation}\label{Hilbert}
    \mathcal{H}=\bigoplus\limits_{\alpha} \mathcal{H}_{\alpha}\;\;\;\;\;\;\;\;\;H_{\alpha}=[\bigotimes\limits_{x\in\mathcal{S}} \mathcal R(\alpha(x))]\otimes M_\alpha
\end{equation}
and $\mathcal{R}(\alpha)$ is the irrep associated to the density $\alpha$ at a point. The Hilbert space $M_\alpha$ carries the muliplicity of that irrep, and is unaffected by the $\widetilde{SL}(2,\mathbb{R})$ action. The diffeomorphism subgroup acts by transforming $\mathcal{H}_\alpha$ according to the transformation of $\alpha$. For an irreducible representation of the whole group, the direct sum must be over an orbit of this action, which is equivalent to summing over all $\alpha$ with fixed integral over $\mathcal{S}$ --- in other words,  the representation is an eigenspace of $\mathcal{C}_1$. Since $\mathcal{C}_1=4\pi$ on the classical phase space, we set this to be the eigenvalue of the quantum representation as well.
\par
Select a representative of this orbit with $\alpha$ equal to some fixed reference density $\tilde{\alpha}$. In order for $G$ to act irreducibly on $\mathcal{H}$, $M_\alpha$ must be an irreducible representation of the little subgroup of the diffeomorphisms that preserves $\tilde{\alpha}$. Since the corner is 1-dimensional, this is just an abelian group of rotations generated by the vector $\xi_{\tilde{\alpha}}$, which has 1-dimensional irreps. Therefore $M_\alpha$ doesn't affect the structure of the Hilbert space in equation \eqref{Hilbert}. The only information it carries is the eigenvalue of the little group generator, which roughly captures the effect that the rotation has on the phase of the state, independent of the action on the $\widetilde{SL}(2,\mathbb{R})$ generators. This appears to correspond to the eigenvalue of the Casimir $\mathcal{C}_2$, though a precise formulation of the connection requires constructing the analogue of the expression \eqref{C_2} in the quantum theory. At any rate, for the purposes of quantizing the original phase space, it appears that this eigenvalue is zero, so the factors $M_\alpha$ can be safely ignored. If it is indeed possible, as suggested at the end of section \ref{sec:coadjoint}, to extend the interpretation of the algebra to BTZ black hole or sourced spacetimes, then this eigenvalue will likely become important.
\par
The properties of the irreducible representation under consideration depend on the class of the unitary representations of $\widetilde{SL}(2,\mathbb{R})$. As found by Bargmann long ago \cite{bargmann}, these come in two families, the discrete series representations that have negative Casimir, and continuous series representations with positive Casimir. For particularly convenient references see \cite{kitaev} for concise formulae and \cite{sl2r_quantization} for an examination of several bases of the algebra, with an emphasis on the physics. The authors of \cite{subsystem} found that the Casimir of their $\mathfrak{sl}(2,\mathbb{R})$ algebra at a point was the area density, so they suggested that the continuous series is physically natural because it guarantees that this is positive definite. In our case, this is not true. Upon rescaling to agree with convention, the corresponding Casimir density is $\frac{\mathcal{C}_{\widetilde{SL}(2,\mathbb{R})}}{\Lambda}$, which is not proportional to the area density. $\mathcal{C}_{\widetilde{SL}(2,\mathbb{R})}$ is always positive on phase space, so in the quantization for negative $\Lambda$, the representation should be in the discrete series, and for positive $\Lambda$ it should be in the continuous series. Meanwhile $\mathcal{Q_B}$ \textit{is} proportional to the area density, but by comparing to the results of \cite{sl2r_quantization} it can be seen that it does not have definite sign in any of the possible representations (which assign all three of the charges to self-adjoint operators).  This fact has a simple classical analog; translations along the horizon can eventually encounter crossings or simply make part of the corner timelike, both of which will locally invert the sign of the area density. These correspond to points on the coadjoint orbit that are not in the classical phase space. Quantum mechanically, however, one can always implement any transformation in $G$, so the inversion of the area density is not avoidable and its spectrum cannot be positive definite. 

\par
Again, this picture is informal due to the infinite dimensionality of the algebra. A precise construction must be in terms of smeared operators. The example of free scalar field theory is instructive; the system can be represented formally as an infinite tensor product of decoupled momentum modes, but the separable Hilbert space that is physically relevant is the completion of a Fock space produced by acting on a selected state with smeared field operators, an example of GNS reconstruction \cite{witten_qft}. In flat spacetime, the preferred choice of state is the vacuum, which is a tensor product of states in each mode Hilbert space. It seems reasonable to attempt a similar construction for the causal diamond quantization, where the analogue of a "momentum mode" is a representation of $\widetilde{SL}(2,\mathbb{R})$ at a single point of the corner. These are decoupled under the dynamics generated by the normal translation and boost generators, which carry the information from the equations of motion. This picture is reminiscent of the black hole horizon algebras discussed in \cite{GSL}.
\par
By analogy, then, the reference state should be some formal superposition over $\alpha$ of states in the tensor product of local mode spaces. Determining expectation values of $\mathcal{\hat{Q}_B}$, $\mathcal{\hat{Q}_+}$, and $\mathcal{\hat{Q}_-}$ in this state will allow the construction of the Fock space and its Hilbert space completion. The problem of defining such a state has two parts. First, one has to choose the states for given values of $\alpha$.  Second, one must choose some measure $\mathcal{D\alpha}$ over the space of densities $\alpha$ with a given total integral over the corner, to allow integration of wavefunctionals of $\alpha$. The unitarity of superrotations demands that this measure be invariant under pullback of $\alpha$. It is only necessary to know the measure up to a multiplication by a function of $\alpha$, since this will only relabel wavefunctionals. Both of these steps will be pursued in further work. 
\par
\textbf{Acknowledgements} I acknowledge  Raman Sundrum, Gong Cheng, and Eanna Flanagan for useful conversations. Special thanks are due to Rodrigo Andrade e Silva and Ted Jacobson, who inspired this project by their prior work and provided invaluable discussion throughout its execution, and to Antony Speranza for critical feedback in its final stages. This work was supported in part by US National Science Foundation grants PHY-2012139 and PHY-2309634.
\bibliography{refs}
\end{document}